\documentclass{PoS}

\usepackage{graphicx}
\usepackage{color}
\pdfoutput=1

\def\simlt{\rlap{\lower 3.5 pt\hbox{$\mathchar \sim$}}\raise 1.0pt \hbox {$<$}}

\title{Hadronic Interactions}

\ShortTitle{Hadronic Interactions}

\author{\speaker{Takeshi Yamazaki}\\%
Faculty of Pure and Applied Sciences,
University of Tsukuba, Tsukuba, Ibaraki 305-8571, Japan\\
Center for Computational Sciences, University of Tsukuba,
Tsukuba, Ibaraki 305-8577, Japan\\
RIKEN Advanced Institute for Computational Science,
Kobe, Hyogo 650-0047, Japan\\
        E-mail: \email{yamazaki@het.ph.tsukuba.ac.jp}}

\abstract{
Understanding hadronic interactions is crucial for investigating 
the properties of unstable hadrons, since measuring physical quantities for 
unstable hadrons including the resonance mass and decay width requires 
simultaneous calculations of  final scattering states.   
Recent studies of hadronic scatterings and decays are reviewed from this point of view.

The nuceon-nucleon and multi-nucleon interactions are very important to 
understand the formation of nucleus from the first principle of QCD. 
These interactions have been studied mainly by two methods,  
due originally to L\"uscher and to HALQCD. 
The results obtained from the two methods are compared in three channels, $I=2$ two-pion, H-dibaryon, and two-nucleon channels.
So far the results from the two methods for the two-nucleon channels 
are different even at the level of the presence or absence of bound states.  
We then discuss possible uncertainties in each method. 

Recent results on the binding energy for helium nuclei are also reviewed.
}

\FullConference{The 32nd International Symposium on Lattice Field Theory\\
		 23-28 June, 2014\\
		 Columbia University New York, NY}

\begin{document}

\section{Introduction}

The ultimate goal of lattice QCD is
to understand the properties of all hadrons quantitatively
by first principle calculations. 
For stable hadrons, recent lattice QCD calculations have reached the level of a precise measurement of physical quantities such as mass and decay constant, thus touching the 
goal. On the other hand, we are far from this goal for unstable hadrons. 
The calculations and analyses involved are much more
difficult and complex, because one has to simultaneously 
treat final scattering states to measure physical quantities of  unstable hadrons.
To calculate physical quantities in the scattering and decay processes,
it is important to understand the interactions between hadrons based on QCD. 

The most famous of hadronic interactions is the nuclear force, which
binds nucleons into nuclei.
Thus, another goal of the lattice QCD is
to prove the formation of nucleus from the degrees of freedom of
quarks and gluons.

For those goals, so far various studies have been carried out.
In this report, recent studies on hadronic scattering, decay,
and light nuclei are reviewed.
Furthermore, it is reviewed consistency and inconsistency
of results obtained from L\"uscher's method and HALQCD methods
in $I=2$ two-pion, H-dibaryon, and two-nucleon channels.
In the two-nucleon channels, possible systematic uncertainties in each
method are discussed.

The organization of this report is as follows.
In the next section, L\"uscher's method is briefly reviewed,
which allows one to evaluate the scattering phase shift of
two particles from finite volume calculations.
In Secs.~\ref{sec:a0} and~\ref{sec:delta}
some recent studies for
the scattering length and physical quantities in decay processes
are presented, respectively.
Sec.~\ref{sec:comparison} presents a  comparison of results from the L\"uscher's method
and HALQCD method, the latter having been recently proposed to calculate the potential
between two hadrons, and a discussion on the uncertainties in each method.
In Sec.~\ref{sec:nuclei} we review recent calculations of light nuclei
and related studies. 
Conclusions are given in Sec~\ref{sec:conclusion}.

\section{L\"uscher's finite volume formula}
\label{sec:luschersmethod}

L\"uscher's formula~\cite{Luscher:1986pf,Luscher:1990ux}
in the original form enables to evaluate the scattering phase shift $\delta(p)$ of a two-particle system in the center-of-mass frame  from the energy of two-particle state
in finite volume, {\it e.g.}, in the S-wave case,
\begin{equation}
\tan\delta(p) = \frac{\pi^{3/2}q}{Z_{00}(1;q^2)},
\end{equation}
where $q = Lp/(2\pi)$ and 
\begin{equation}
Z_{00}(s;q^2) = \frac{1}{\sqrt{4\pi}}
\sum_{{\vec n}\in Z^3}\frac{1}{\left({\vec n}^2 - q^2\right)^s} .
\end{equation}
The relative momentum between the two particles, $p$, is determined from
the two-particle energy, $W^2 = 4( m^2 + p^2 )$ with
$m$ being mass of the particle.
An important assumption is that
the interaction of the two particles does not depend on volume, 
and that the interaction is well localized, {\it i.e.}, 
there is a region such that $V(\vec{r}) \approx 0$ for 
$r < L/2$ with $L$ being the linear spatial extent.

An extension of L\"uscher's formula to moving frame,
where the total momentum of two particles is non-zero,
was first proposed by Rummukainen and Gottlieb~\cite{Rummukainen:1995vs}.
Their result was confirmed by different 
derivations~\cite{Kim:2005gf,Christ:2005gi},
and extensions in moving frame with different total momenta,
and several irreducible representations
have also been derived~\cite{Feng:2011ah,Dudek:2012gj}.
In the recent papers~\cite{Fu:2011xz,Leskovec:2012gb,Doring:2012eu,Gockeler:2012yj,Li:2012bi} extensions of the formula in moving frames to the case of particles with different masses 
are derived, in which case mixings occur between even and odd partial waves.
Those extensions are also reviewed by other plenary speakers~\cite{Briceno:2014pka,Prelovsek:2014zga}.

L\"uscher's formula encompasses the case of a bound state as a pure imaginary solution for the relative momenta $p$, and indicates that the size dependence of the energy shift of the ground two-particle state relative to the free two-particle state is exponentially small~\cite{Beane:2003da,Sasaki:2006jn}. 
With this result as a background, study of size dependence of the energy shift has been employed in a number of recent papers to examine bound state formation in multi-hadron systems~\cite{Beane:2010hg,Yamazaki:2011nd,Yamazaki:2012hi,Beane:2012vq,Yamazaki:2015asa}. 
We shall call the methodology in these studies generally as L\"uscher's method.

\section{Scattering length}
\label{sec:a0}

The scattering length $a_0$ is the simplest physical quantity 
in an S-wave scattering process, which is defined from the phase shift $\delta(p)$
by $a_0 = \lim_{p\to 0}[\tan\delta(p)/p]$ in the zero relative momentum limit.

\subsection{$I=2$ two-pion channel}

The $I=2$ two-pion scattering is a very simple process involving 
no bound states or resonances in the low energy region.
It is also the simplest from lattice calculation point of view, 
where only the direct and cross diagrams are necessary for the two-pion
correlation function.
While there have been various quenched calculations of $a_0$
in this channel,
only dynamical calculations are reviewed in this report.

\begin{figure}[htbp]
\centering
\includegraphics[width=7cm,clip]{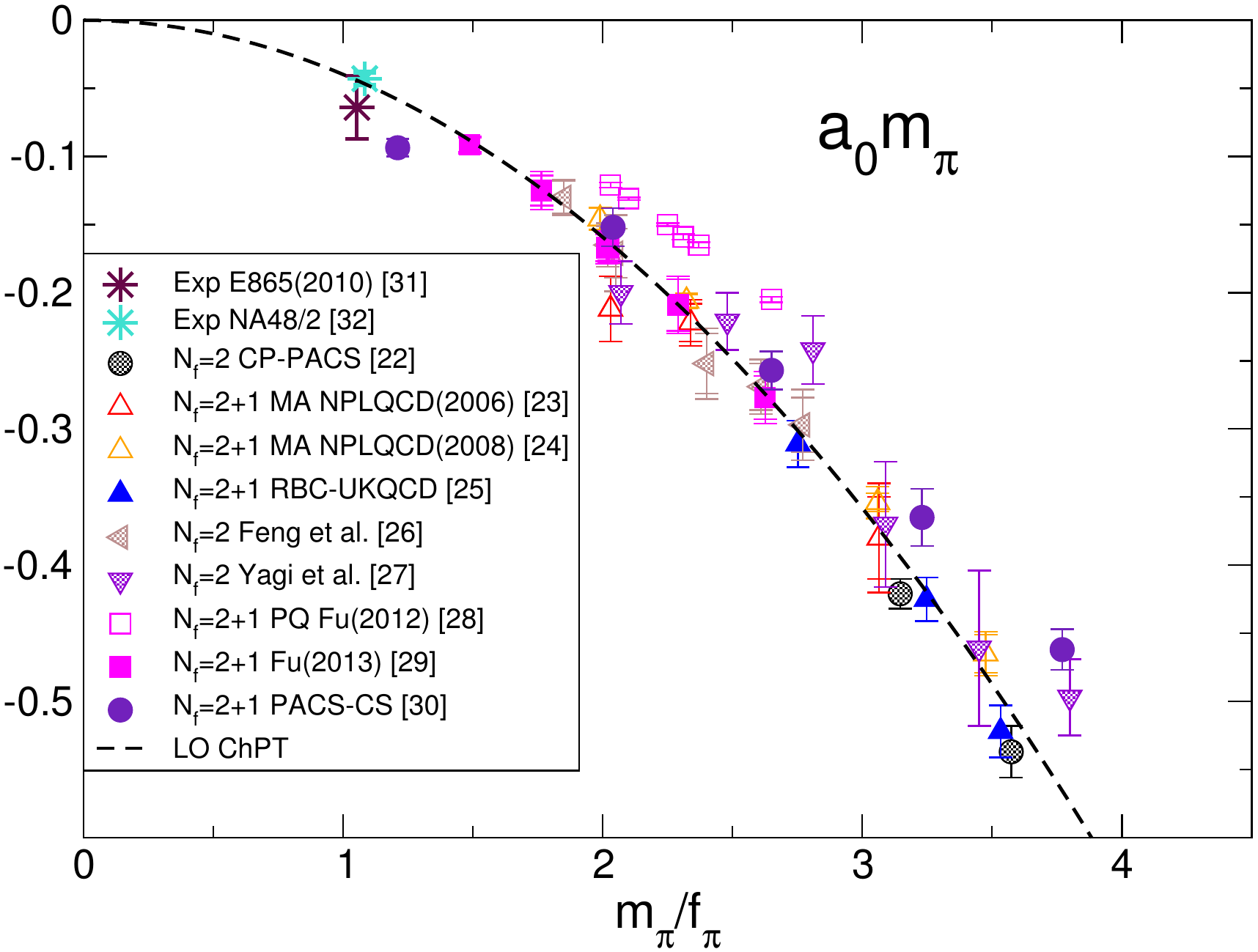}
\includegraphics[width=7cm,clip]{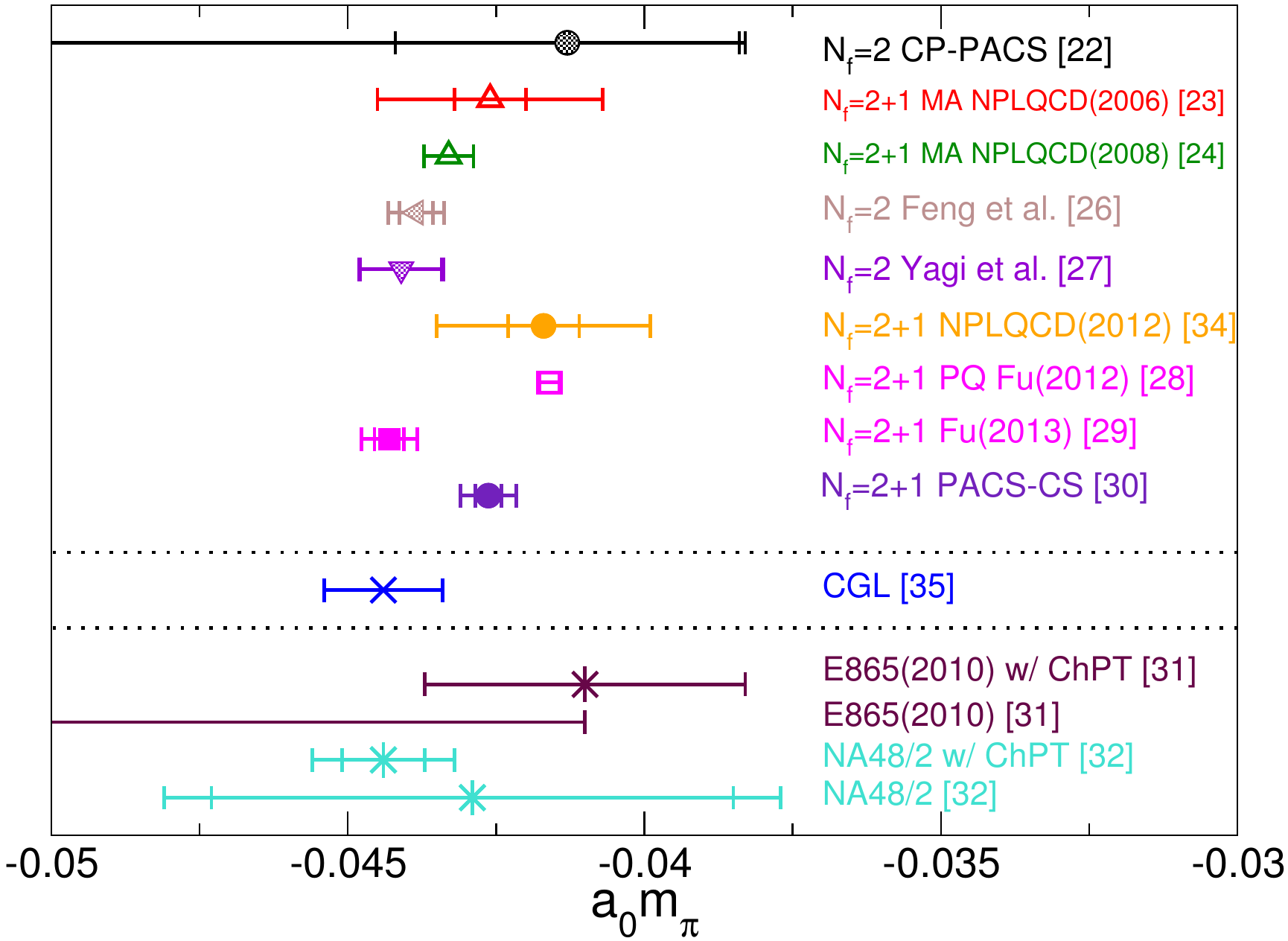}
\caption{
Left panel: $a_0$ in the $I=2$ two-pion channel calculated in full lattice QCD 
as a function of $m_\pi/f_\pi$, together with the prediction of LO ChPT 
(dashed curve). 
Right panel: values at physical $m_\pi$. 
Circle, up triangle, left triangle, down triangle, and square
symbols denote results obtained with clover, domain wall, twisted mass, 
overlap, and improved staggered
quark actions for valence quark, respectively.
MA and PQ express mixed action with asqtad sea quark and 
partially quenched calculation.
The experimental result by
E865~\cite{Pislak:2003sv} and NA48/2~\cite{Batley:2010zza},
and the phenomenological
determination using NNLO ChPT~\cite{Colangelo:2001df} (denoted by CGL in 
the right panel) are also shown.
}
\label{fig:I2_pipi}
\end{figure}

The left panel in Fig.~\ref{fig:I2_pipi} shows 
the results for $a_0 m_\pi$ 
from dynamical calculations~\cite{Yamazaki:2004qb,Beane:2005rj,Beane:2007xs,Liu:2009uw,Feng:2009ij,Yagi:2011jn,Fu:2011bz,Fu:2013ffa,Sasaki:2013vxa}
together with two experimental results
by E865~\cite{Pislak:2003sv} and NA48/2~\cite{Batley:2010zza} plotted with 
burst symbols.
The calculations by Hadron Spectrum~\cite{Dudek:2010ew,Dudek:2012gj} 
and NPLQCD~\cite{Beane:2011sc} Collaborations 
in $N_f = 2+1$ full QCD are not plotted here,
because the pion decay constant has not been calculated in the simulation
parameter.
The lattice results are compared with LO (leading order) ChPT
(chiral perturbation theory) formula,
$a_0 m_\pi = -m_\pi^2/(8\pi f_\pi^2)$, drawn by the dashed curve, 
where $f_\pi$ is the pion decay constant with 
the normalization of $f_\pi = 132$ MeV at the physical pion mass.
Most of lattice results agree with the formula. 
In the small $m_\pi/f_\pi$ region, however,  
partially quenched staggered (open square)~\cite{Fu:2011bz}
and nonperturbatively improved Wilson (closed circle)~\cite{Sasaki:2013vxa}
results significantly deviate from LO ChPT.

The reason for the deviation of the staggered calculation~\cite{Fu:2011bz} 
is not clear. 
The lightest data in Ref.~\cite{Fu:2011bz} (open square) 
should agree with the one at the same $m_\pi/f_\pi$ in Ref.~\cite{Fu:2013ffa}
(closed square), because
their actions and simulation parameters are the same.
On the other hand, the deviation of the improved Wilson calculation 
is probably due to chiral symmetry breaking effect of the Wilson quark action 
at non-zero lattice spacings~\cite{Sasaki:2013vxa}.

The lattice results for $a_0$ at the physical $m_\pi$ 
are shown in the right panel of Fig.~\ref{fig:I2_pipi}.
For the chiral extrapolation of $a_0$,
NLO (next-to-leading order) ChPT formula
is employed in 
almost all calculations~\cite{Beane:2005rj,Feng:2009ij,Fu:2011bz,Fu:2013ffa}.
On the other hand, 
Refs.~\cite{Beane:2007xs,Sasaki:2013vxa} use
variations of NLO ChPT formula 
including systematic errors coming from finite lattice spacings.
The value of Ref.~\cite{Beane:2011sc} is estimated using 
the low energy constants extracted from $\delta(p)$ at
$m_\pi = 390$ MeV with NLO ChPT formula for the scattering amplitude,
while they use the value of $f_\pi$ calculated with different action.

The lattice results are compared with the phenomenological
determination, marked ``CGL'', obtained with NNLO ChPT~\cite{Colangelo:2001df} 
and the experimental results~\cite{Batley:2010zza,Pislak:2003sv}
with and without ChPT constraint.
Since in ChPT the scattering lengths for the $I=0$ and 2 channels share
a common low energy constant, the error becomes smaller in the case with
the constraint.
Some lattice results have smaller error, 
even if including the systematic error, than the
phenomenological determination and the experimental results.
Therefore, $a_0$ calculation in the $I=2$ two-pion channel
enters precision measurement era, where lattice result
will be confirmed by experiment.
In this conference, preliminary result of $N_f = 2+1+1$
twisted mass quark calculation~\cite{Helmes:2014wca} is reported, 
which will calculate more precise value of $a_0$ at the physical $m_\pi$.

\subsection{$I=1/2$ $K\pi$ channel}

The S-wave scattering in the $I=1/2$ $K\pi$ channel
is more complicated than the $I=2$ two-pion channel; it is in fact much more interesting 
because of the expected presence of $\kappa$ meson from phenomenology 
in this channel. 
In the region of heavy quark mass, $\kappa$ meson is indeed a bound state 
and so has to be taken into account in the $a_0$ calculation of $K\pi$
scattering~\cite{Sasaki:2013vxa}. 
In the case, one needs to calculate
the first excited state in the system to evaluate $a_0$.
Furthermore, for the calculation of $K\pi$ correlation function,
we need to include the rectangular diagram, 
for which specific calculation methods are required,
for example the stochastic LapH method~\cite{Morningstar:2011ka}
or the  trapezoid diagram calculation method~\cite{Aoki:2011yj}.

\begin{figure}[htbp]
\centering
\includegraphics[width=7cm,clip]{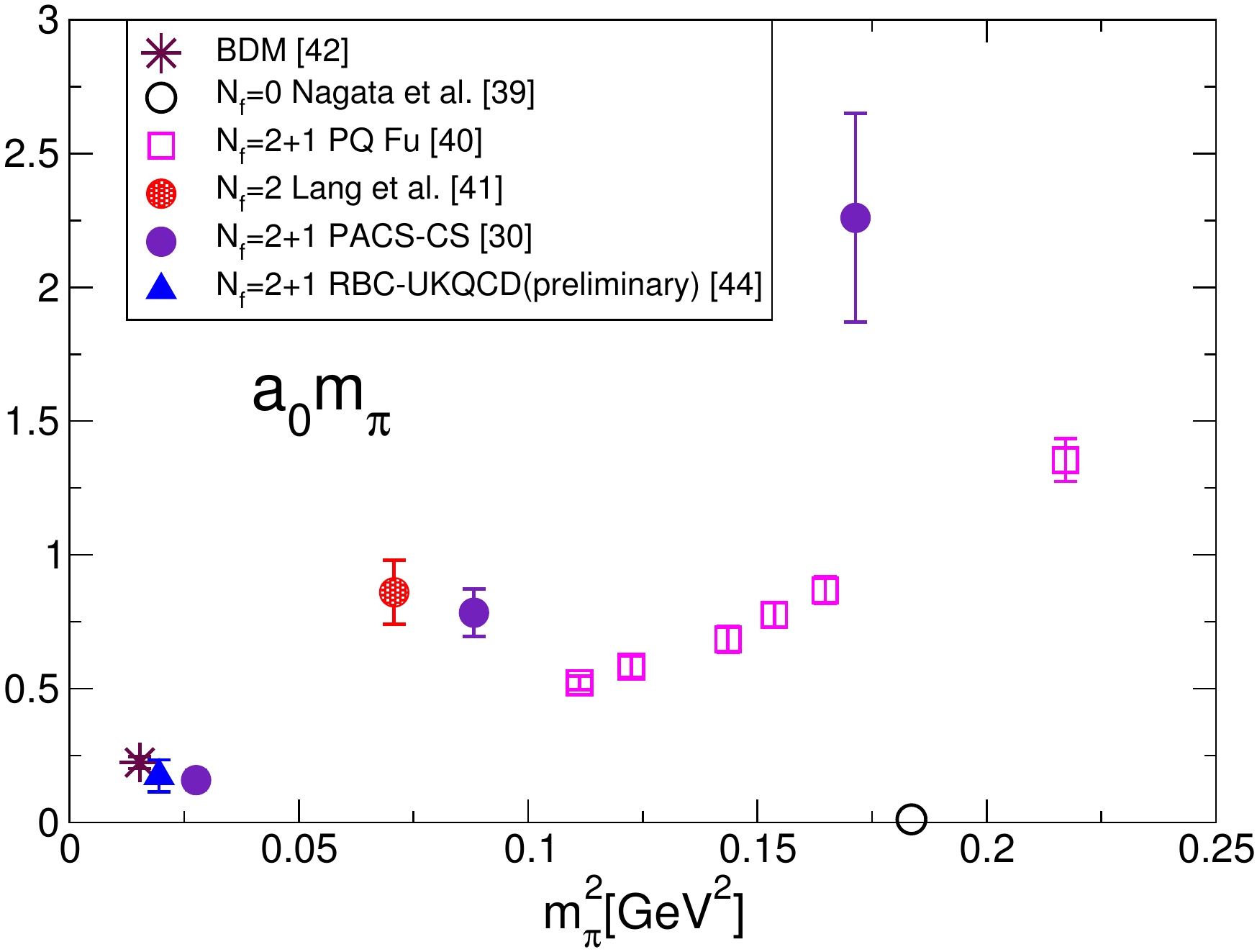}
\includegraphics[width=7cm,clip]{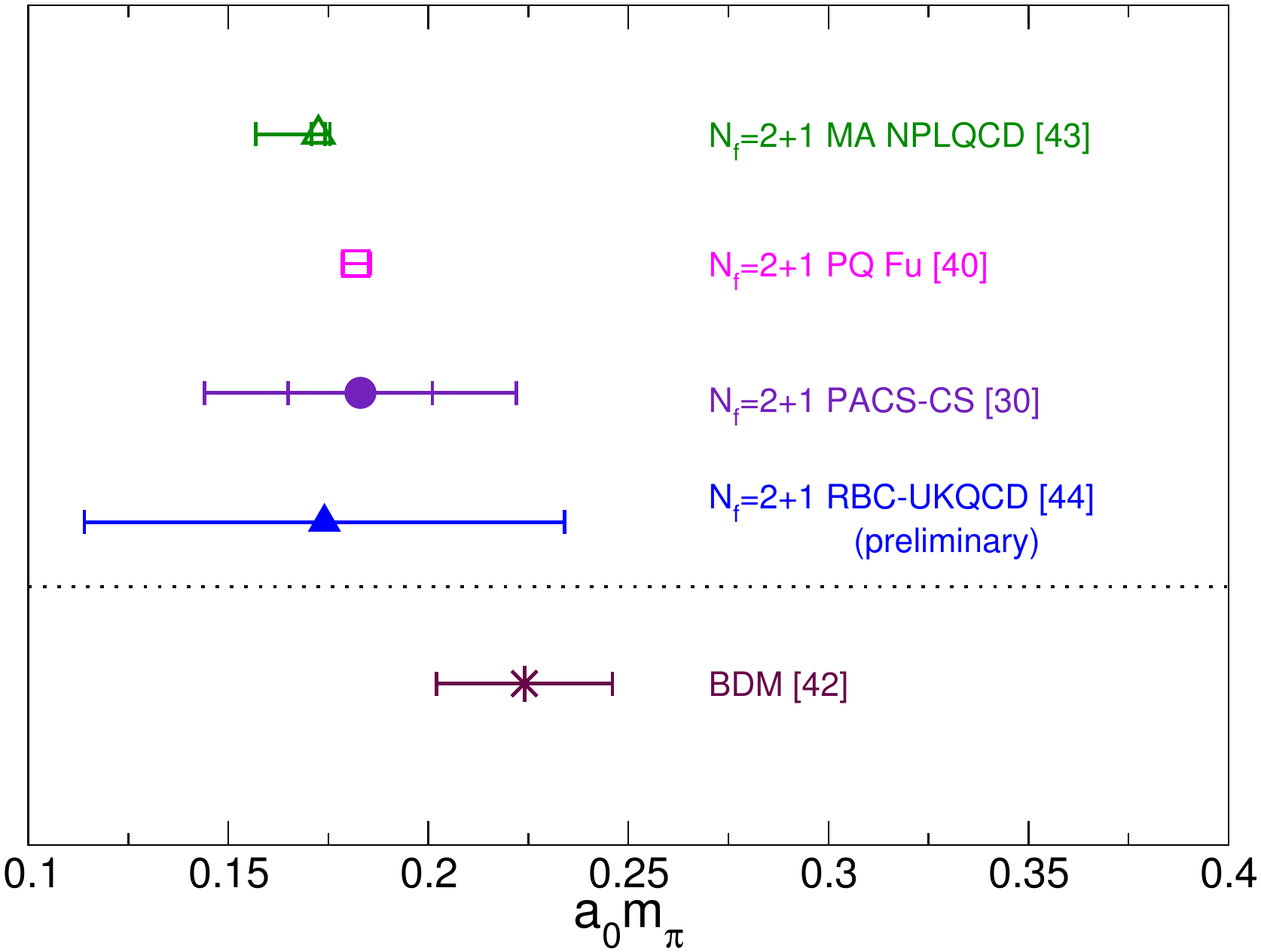}
\caption{
Left panel: $a_0 m_\pi$ as a function of $m_\pi^2$ in the $I=1/2$ $K\pi$ channel.  
Right panel: $a_0 m_\pi$ at physical $m_\pi$.  
Phenomenological determination
by B\"uttiker {\it et al.}~\cite{Buettiker:2003pp}, 
and an indirect determination by NPLQCD Collaboration~\cite{Beane:2006gj} 
are also shown.
Burst symbol in left panel is shifted for clarify.
}
\label{fig:I12_kpi}
\end{figure}

Due to these difficulties,
there have been only four studies in
the $I=1/2$ $K\pi$ channel; 
$N_f = 0$ calculation by Nagata {\it et al.}~\cite{Nagata:2008wk},
partially quenched $N_f = 2+1$ by Fu~\cite{Fu:2011wc}, 
$N_f = 2$ by Lang {\it et al.}~\cite{Lang:2012sv}, and
$N_f = 2+1$ by PACS-CS Collaboration~\cite{Sasaki:2013vxa}.
The left panel of Fig.~\ref{fig:I12_kpi} shows the result
of the four studies for $a_0 m_\pi$, together with 
phenomenological determination~\cite{Buettiker:2003pp}.
The lattice data do not show
a common $m_\pi^2$ dependence in contrast to the situation with 
the $I=2$ two-pion channel in Fig.~\ref{fig:I2_pipi}.
This is presumably due to the systematic error inherent in each calculations, 
{\it e.g.,}
a small spatial size of $\sim 2$ fm in the calculation 
of Ref.~\cite{Lang:2012sv}
and chiral symmetry breaking effect at the lightest data point  
of Ref.~\cite{Sasaki:2013vxa}.

Similar to the calculations with the $I=2$ two-pion channel,
$a_0 m_\pi$ at the physical quark mass is estimated using
NLO ChPT formula in Ref.~\cite{Fu:2011wc}
and its extension with chiral symmetry breaking term 
in Ref.~\cite{Sasaki:2013vxa}.
These results are plotted in the right panel of Fig.~\ref{fig:I12_kpi}.
The panel also presents an indirect estimation by 
NPLQCD Collaboration~\cite{Beane:2006gj}, which uses 
the result of $a_0$ in the $I=3/2$ $K\pi$ channel,
since $a_0$ for $I=1/2$ and $I=3/2$ $K\pi$ channels share
the common low energy constants in NLO ChPT formulae.

Lattice results are in rough accord with the phenomenological determination. 
While this is encouraging, in order to make more precise comparison with
experiment,
further direct calculations would be desired.
In this conference, as shown in Fig.~\ref{fig:I12_kpi},
RBC-UKQCD Collaboration~\cite{Janowski:2014lat} reports
preliminary result of direct calculation of $a_0$ at the physical quark masses,
where chiral extrapolation is not needed, in a large spatial extent
of 5.5 fm.
Furthermore, after the conference, Hadron Spectrum 
Collaboration presents the result at $m_\pi = 390$ MeV~\cite{Wilson:2014cna}.

\section{Resonances}
\label{sec:delta}

It is possible to obtain the mass and decay width of resonances  
from the phase shift $\delta(p)$ of its final scattering state, at least in
two-particle decay processes.
For the $\rho$ meson decay into two pions, for example,
the $\rho$ meson mass $m_\rho$ and an effective $\rho\pi\pi$ coupling $g_{\pi\pi\rho}$ 
can be extracted through the Breit-Wigner form, 
\begin{equation}
\frac{p^3}{\sqrt{s}}\cot \delta(p) = 
\frac{6\pi}{g_{\pi\pi\rho}^2}(m_\rho^2 - s),
\label{eq:B-W_form}
\end{equation}
where $\sqrt{s}$ is the center-of-mass energy of the two pions.
The decay width $\Gamma_\rho$ is then obtained by 
\begin{equation}
\Gamma_\rho = \frac{p^3_{\rho}}{m_\rho^2}\frac{g^2_{\rho\pi\pi}}{6\pi},
\end{equation}
where $p_{\rho}$ is the relative momentum of the two pions at the 
resonance point ($s = m_\rho^2$),
$p_{\rho}^2 = m_\rho^2/4-m_\pi^2$.
Therefore, when $\delta(p)$ is calculated by
L\"uscher's method or its extensions for at least two different values of $p$,
$m_\rho$ and $g_{\pi\pi\rho}$ (or $\Gamma_\rho$) can be determined 
using eq.(\ref{eq:B-W_form}).

In order to obtain the phase shift for several different values of $p$,
we might need to extract the excited state energy corresponding to 
higher scattering states.
For this purpose,
the variational method
proposed by L\"uscher and Wolff~\cite{Luscher:1990ck}
is often used in lattice calculations.

\subsection{$I=1$ two-pion channel}

The decay process of the $\rho$ meson is an ideal resonance process for an initial study of scattering and decay,
since the spectrum consists only of the $\rho$ meson 
and its final two-pion scattering state
from the threshold ($2m_\pi$) to the vicinity of the resonance energy region. 
Computationally, however, this decay process is 
much harder than the $a_0$ calculation of the $I=2$ two-pion channel.
One of the difficulties is the necessity of the rectangular diagram, 
as in the $I=1/2$ $K\pi$ channel,
with a non-zero relative momentum of the two pions.
Another difficulty is that, since the decay proceeds through the P-wave, 
the pion has to be sufficiently light and the spatial size $L$ sufficiently 
large so as to satisfy the decay condition,
\begin{equation}
m_\rho^2 > 4\left(m_\pi^2 + (2\pi/L)^2\right).
\end{equation}
This condition can be relaxed if one uses the moving 
frame~\cite{Rummukainen:1995vs}.

\begin{figure}[htbp]
\centering
\includegraphics*[width=7cm,clip]{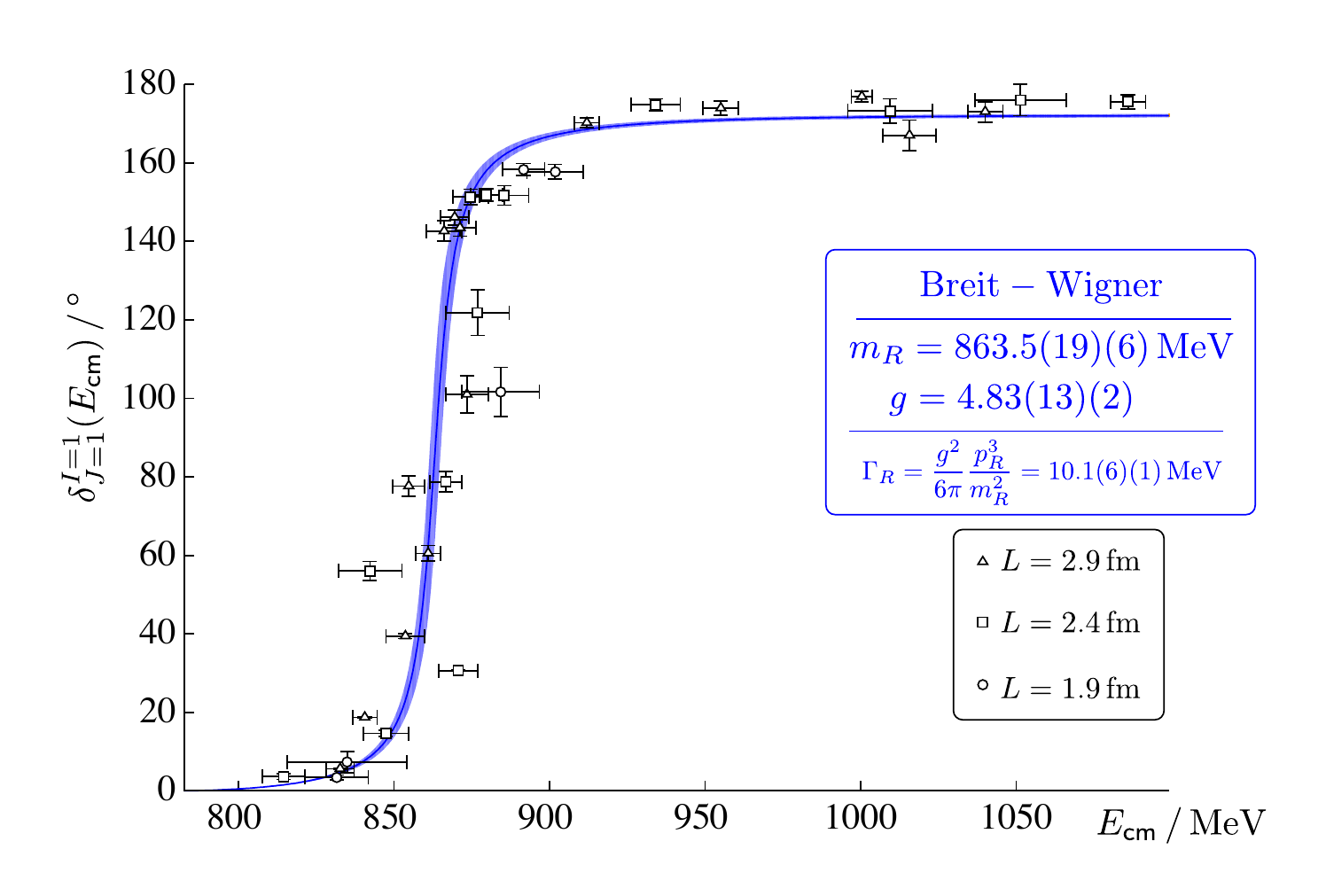}
\caption{$I=1$ $\pi \pi $ phase shift calculated by Hadron Spectrum Collaboration~\cite{Dudek:2012xn}.
}
\label{fig:rho_delta}
\end{figure}

The first calculation of the $\rho$ meson decay was carried out
in $N_f = 2$ QCD by CP-PACS Collaboration~\cite{Aoki:2007rd} using
the total momentum $|P| = 1$ frame.
After this work, five papers~\cite{Feng:2010es,Lang:2011mn,Aoki:2011yj,Pelissier:2012pi,Dudek:2012xn} have been published.
Figure~\ref{fig:rho_delta} show $\delta(p)$
calculated with $m_\pi = 390$ MeV by 
Hadron Spectrum Collaboration~\cite{Dudek:2012xn}.
There are 29 data points obtained at different $p$'s  using several moving frames and
irreducible representations on three different volumes.
From the Breit-Wigner  fit as in eq.~(\ref{eq:B-W_form}),
they extract the resonance mass $m_R = m_\rho$ and 
the coupling $g_{\pi\pi\rho}$ as shown in the figure.

\begin{figure}[htbp]
\centering
\includegraphics[width=7cm,clip]{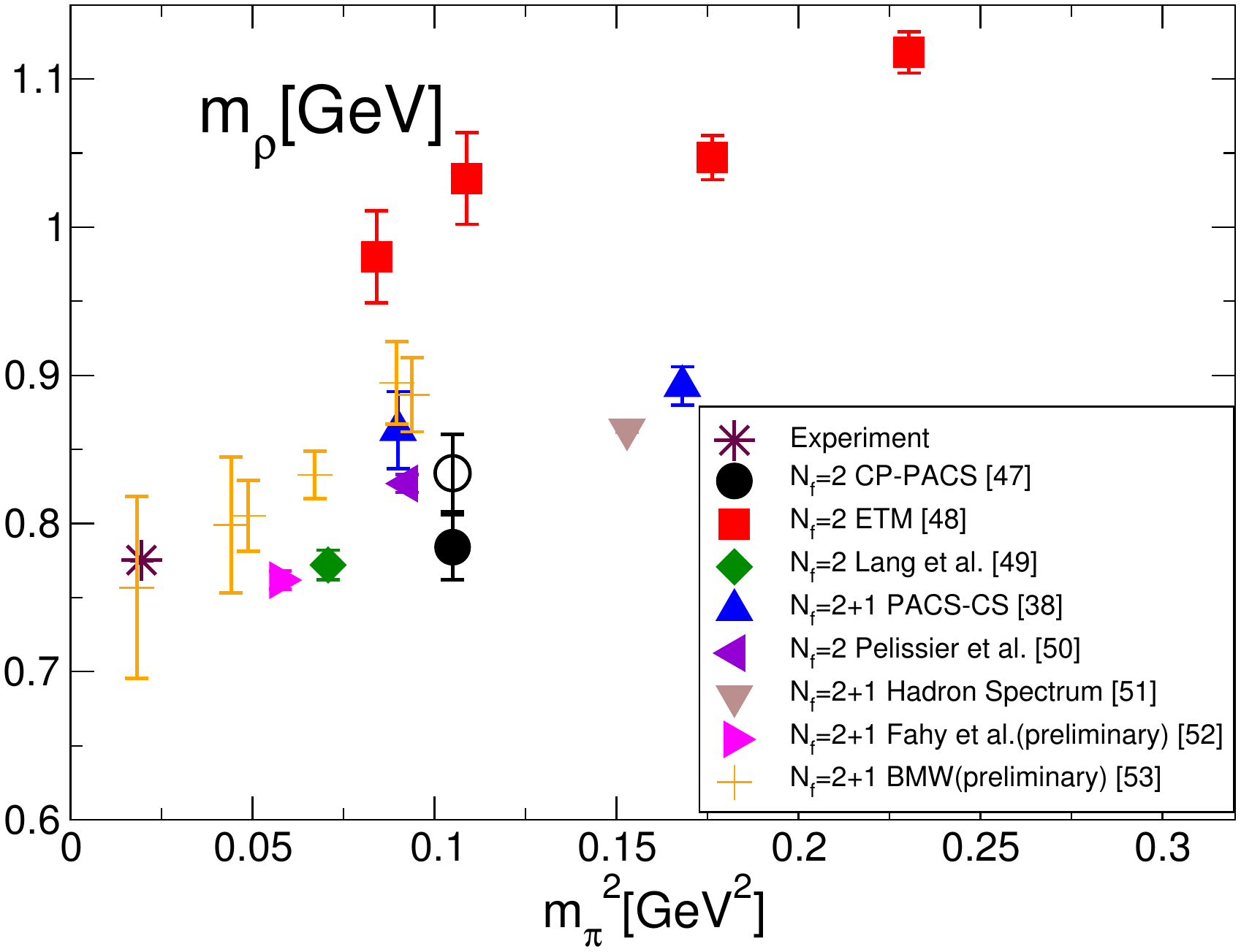}
\includegraphics[width=7cm,clip]{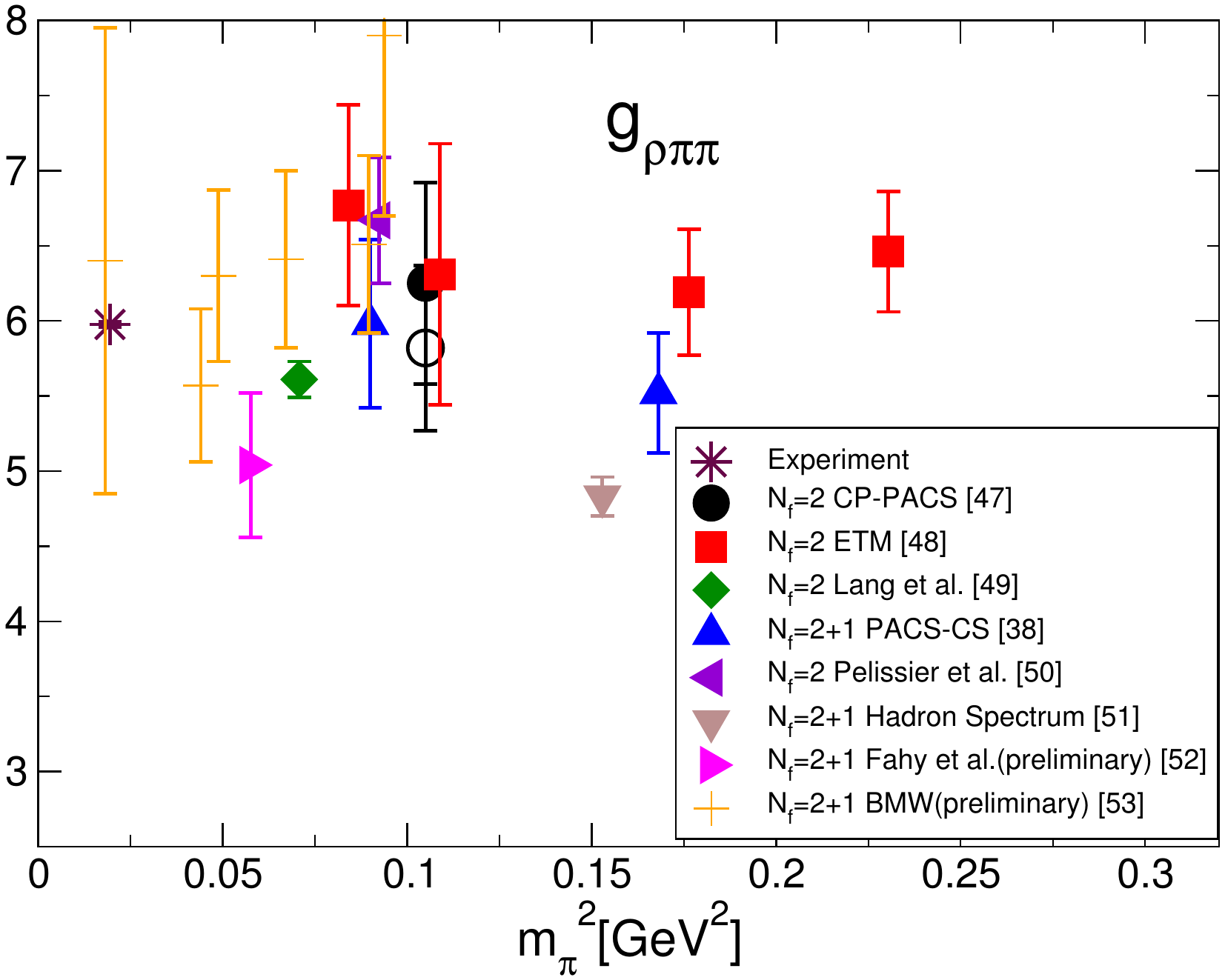}
\caption{
Resonance mass $m_\rho$ (left panel) and the coupling $g_{\pi\pi\rho}$ (right panel) obtained in various calculations.
Open circle is result with lattice dispersion relation~\cite{Aoki:2007rd}.
}
\label{fig:rho}
\end{figure}

Figure~\ref{fig:rho} presents a compilation of lattice results for
$m_\rho$ (left panel) and $g_{\pi\pi\rho}$ (right panel)
in Refs.~\cite{Aoki:2007rd,Feng:2010es,Lang:2011mn,Aoki:2011yj,Pelissier:2012pi,Dudek:2012xn}
extracted from the Breit-Wigner form fit.
The experimental values are also plotted.
For $m_\rho$ result, the $N_f = 2$ results by ETM Collaboration
are larger than the others.  
This discrepancy could be explained by a systematic error
from scale setting.
If we choose Sommer scale $r_0$ for the scale setting in all the results,
discrepancy becomes smaller, for example, the results for ETM and PACS-CS Collaborations differ by less than 7\%~\cite{Aoki:2011yj}.

The lattice results for $g_{\pi\pi\rho}$, presented in the right panel
of Fig.~\ref{fig:rho}, show that it has only weak dependence on the pion mass, 
and that its value is consistent with experiment 
albeit errors are still generally large. 
We should note that the result by Hadron Spectrum Collaboration~\cite{Dudek:2012xn} with a small error does not quite agree with experiment, 
and so is that by Lang {\it et al.}~\cite{Lang:2011mn} 
albeit with a better agreement.  
The discrepancies from the experiment are not decreased by
choosing $r_0$ to set the scale.
In this conference, two preliminary results
by Fahy {\it et al.}~\cite{Fahy:2014jxa}
and BMW Collaboration~\cite{Metivet:2014bga}
are reported as shown in Fig.~\ref{fig:rho}.
Fahy {\it et al.}~\cite{Fahy:2014jxa} obtain
the result at $m_\pi = 240$ MeV, and
BMW Collaboration obtain the results in 
$m_\pi = 134$--300 MeV~\cite{Metivet:2014bga}.

\subsection{Other decay channels}

In the P-wave $I=1/2$ channel,
the $K^*$ meson decays to the $K\pi$ final state. 
This is an ideal decay process as in the $\rho$ meson decay; 
there are only $K\pi$ scattering and $K^*$ state in
the energy region below the resonance.
There have been three lattice calculations of $\delta(p)$
in this channel, by Fu and Fu~\cite{Fu:2012tj} in $N_f = 2+1$ QCD,
Prelovsek {\it et al.}~\cite{Prelovsek:2013ela} in $N_f = 2$ QCD,
and Hadron Spectrum Collaboration~\cite{Dudek:2014qha,Wilson:2014cna}
in $N_f = 2+1$ QCD.
The last one is reported in this conference, and 
reviewed by other plenary speakers~\cite{Briceno:2014pka,Prelovsek:2014zga}.
A difficulty of the lattice calculation in this channel
is mixing of even and odd partial wave contributions
in moving frame~\cite{Fu:2011xz,Leskovec:2012gb,Doring:2012eu,Gockeler:2012yj,Li:2012bi}
due to $m_\pi \ne m_K$.
Some methods to solve the difficulty were proposed in
Refs.~\cite{Ozaki:2012ce,Prelovsek:2013ela,Wilson:2014cna}.

In Ref.~\cite{Fu:2012tj} it is assumed that S-wave contribution is negligible,
so that the result contains systematic errors from the mixing.
Prelovsek {\it et al.}~\cite{Prelovsek:2013ela} 
obtain a value consistent with the experiment,
$g_{\pi K K^*}=5.7(1.6)$, at $m_\pi = 270$ MeV.
Hadron Spectrum Collaboration~\cite{Dudek:2014qha,Wilson:2014cna} 
calculate $\delta(p)$ at $m_\pi = 390$ MeV, while
the $K^*$ meson is a bound state at this value of $m_\pi$. 
Hadron Spectrum Collaboration also carry out coupled channel analyses 
in S- and D-wave
channels to extract $\delta(p)$ for each of $\pi K$ and $\eta K$
scatterings and inelasticity.

We should add that studies related to the decay channels,
$N^*\to N\pi$~\cite{Verduci:2014csa}, 
$\sigma\to\pi\pi$~\cite{Giedt:2014ysa,Wakayama:2014gpa}, and
$a_0\to\pi\eta$~\cite{Abdel-Rehim:2014zwa}
are reported at this conference.

\section{Comparisons of L\"uscher's and HALQCD methods}
\label{sec:comparison}

In this section, recent results obtained by 
L\"uscher's method and HALQCD method are compared in three channels,
$I=2$ two-pion channel, H-dibaryon ($\Lambda\Lambda$) channel,
and two-nucleon $NN$ channel.
After the comparison, possible uncertainties 
in each of the two methods are discussed.

\subsection{HALQCD method}

The HALQCD method~\cite{Ishii:2006ec,Aoki:2009ji} extracts 
the potential between two particles, $V(\vec{r})$, from
Nambu-Bethe-Salpeter (NBS) wave function.
The idea derives from the derivation of L\"uscher's formula in
quantum mechanics~\cite{Luscher:1990ux} and
in quantum field theory~\cite{Lin:2001ek}, and also 
from calculations of the NBS wave function in two-dimensional scalar
field theory~\cite{Balog:1999ww} and $I=2$ two-pion channel~\cite{Aoki:2005uf}.
While these studies concentrated on the behavior of NBS wave function at large $r$, where $V(\vec{r})\approx 0$,
the HALQCD method asserts that dynamics could be extracted from the smaller $r$ region.

The NBS wave function of the $n$-th $NN$ state, $\phi_n(\vec{r})$,
in $NN$ channel is calculated from
four-point function of $N$ operator,
\begin{equation}
C_{NN}({\vec r}, t) = 
\sum_{\vec x}\langle 0 | N({\vec x}+{\vec r},t)N({\vec x},t)
\overline{NN}(0)|0\rangle = \sum_n A_n \phi_n({\vec r}) e^{-W_nt}
+ \cdots,
\label{eq:fourpt}
\end{equation}
where $\overline{NN}$ is a $NN$ source operator,
$W_n$ is the $n$-th energy, $W_n = 2 \sqrt{m_N^2 + p_n^2}$,
$A_n = \langle NN,W_n | \overline{NN} | 0 \rangle$,
and 
\begin{equation}
\phi_n(\vec{r}) = 
\sum_{\vec x}\langle 0 | N({\vec x}+{\vec r})N({\vec x})|NN,W_n\rangle .
\end{equation}
The spin index of the $N$ operator is omitted for simplicity.
The dots represent contributions from inelastic scattering states.
A non-local potential $U({\vec r},{\vec r}^\prime)$
is defined by acting with the Laplacian on $\phi_n(\vec{r})$,
\begin{equation}
\left(\frac{\nabla^2}{m_N} + \frac{p_n^2}{m_N}\right)\phi_n({\vec r}) = 
\int d^3r^\prime U({\vec r},{\vec r}^\prime) \phi_n({\vec r}^\prime).
\label{eq:nlp}
\end{equation}
Assuming that the higher order terms of the velocity expansion
of $U({\vec r},{\vec r}^\prime)$ are negligible,
\begin{equation}
U({\vec r},{\vec r}^\prime) \approx V(\vec{r})\delta(\vec{r}-\vec{r}^\prime),
\label{eq:assump_v}
\end{equation}
a local potential $V(\vec{r})$ is obtained from eq.(\ref{eq:nlp}).
It should be noted that in quantum field theory
the right-hand side of eq.(\ref{eq:nlp})
corresponds to 
Fourier transformation of off-shell scattering amplitude~\cite{Aoki:2005uf}, 
which is related to $\delta(p)$ only on shell.

One way to calculate $V(\vec{r})$ is to use the ground state NBS
wave function $\phi_0({\vec r})$ extracted from $C_{NN}({\vec r}, t)$
for large $t$.  In the $NN$ case, however, statistical noises increase exponentially at large $t$, rendering an accurate calculation of $C_{NN}({\vec r}, t)$ difficult. 
It was proposed in Ref.~\cite{HALQCD:2012aa}  
to employ the time dependence
of $C_{NN}({\vec r}, t)$ and rewrite eq.~(\ref{eq:nlp}) as, 
\begin{equation}
\left(\frac{\nabla^2}{m_N}
+ \frac{1}{4m_N}\frac{\partial^2}{\partial t^2}
- \frac{\partial}{\partial t}\right)
\overline{C}_{NN}({\vec r}, t)
= V(\vec{r})\overline{C}_{NN}({\vec r}, t),
\label{eq:tdep_v}
\end{equation}
where $\overline{C}_{NN}(\vec{r},t) = C_{NN}({\vec r}, t) \exp(2m_Nt)$.
A key assumption for this equation is that 
all $\phi_n(\vec{r})$'s in $C_{NN}({\vec r}, t)$
yield a common $V(\vec{r})$, in other words,
eq.(\ref{eq:assump_v}) is valid.
With eq.(\ref{eq:tdep_v}), $V(\vec{r})$ can be obtained from
$C_{NN}({\vec r}, t)$ at smaller $t$ region as long as contributions of inelastic states are negligible.

In the HALQCD method, thus, 
the potential $V(\vec{r})$ obtained is fitted by a continuous function of $r$,
like the Yukawa potential plus some hard core,
which is then employed to calculate the phase shift $\delta(p)$ by solving the Schr\"odinger equation.  
An assumption in calculating $\delta(p)$ is that
finite volume effects are negligible in $V(\vec{r})$.  It is also assumed that, 
while the potential $V(\vec{r})$ may depend on the choice of the sink operator 
employed for $C_{NN}({\vec r}, t)$ in eq.(\ref{eq:fourpt}),
such a dependence does not affect the phase shift $\delta(p)$ evaluated 
from the potential.

\begin{figure}[htbp]
\centering
\includegraphics[width=7cm,clip]{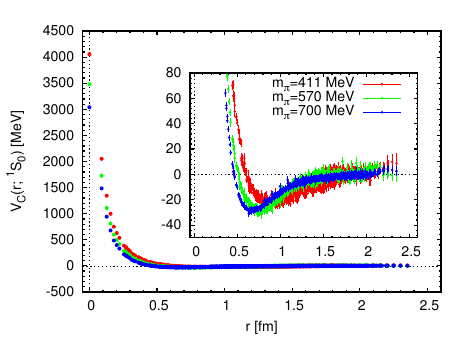}
\raisebox{5mm}{
\includegraphics[width=7cm,clip]{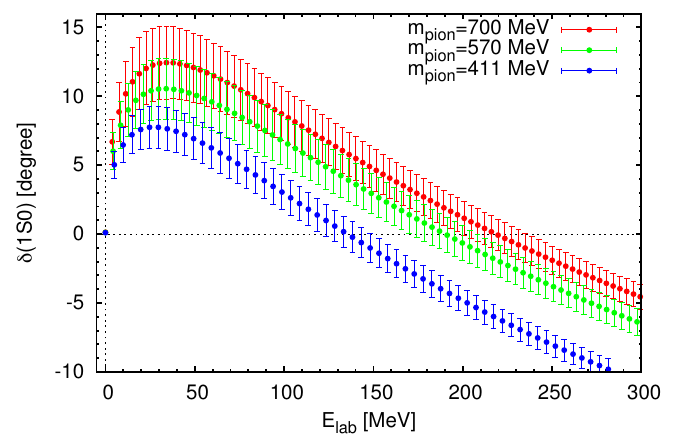}
}
\caption{
Recent result of nuclear potential (left panel) 
and $\delta(p)$ obtained from the potential (right panel)
in $N_f = 2+1$ QCD calculated by HALQCD Collaboration~\cite{Ishii:2013ira}.
}
\label{fig:nf2+1:nforce}
\end{figure}

In the left panel of Fig.~\ref{fig:nf2+1:nforce} we present a recent result 
of HALQCD for the nuclear potential of the spin-singlet $^1$S$_0$ channel
calculated in $N_f = 2+1$ QCD 
at $m_\pi = 410$--700 MeV~\cite{Ishii:2013ira}.
The potential has the features similar to 
the phenomenological nuclear potentials 
with a repulsive core in the smaller $r$ region and 
a one-pion exchange potential at large $r$. 
The depth of the lattice potential seems shallower than the 
phenomenological ones, which is considered to be caused by a heavy $m_\pi$ 
used in the calculation. 
The right panel of Fig.~\ref{fig:nf2+1:nforce} 
presents the results of $\delta(p)$ at each $m_\pi$
calculated from $V(r)$.

In this conference, using the HALQCD method,
an update of the $\Omega\Omega$ potential~\cite{Yamada:2014jra} and
a study of the quark mass dependence of the three-nucleon potential~\cite{Doi:2011gq}
are presented by HALQCD Collaboration.

\subsection{$I=2$ two-pion scattering}

\begin{figure}[htbp]
\includegraphics[width=7cm,clip]{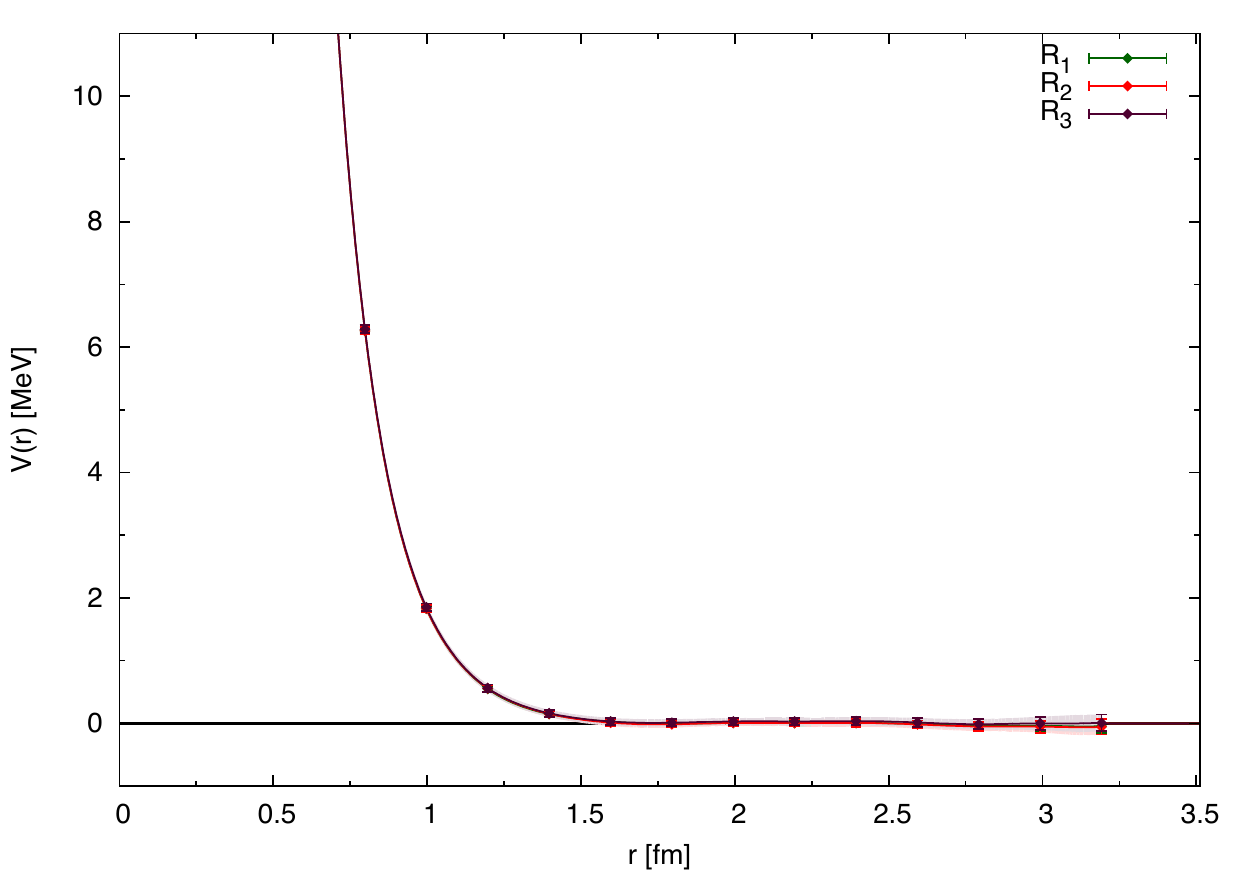}
\includegraphics[width=7cm,clip]{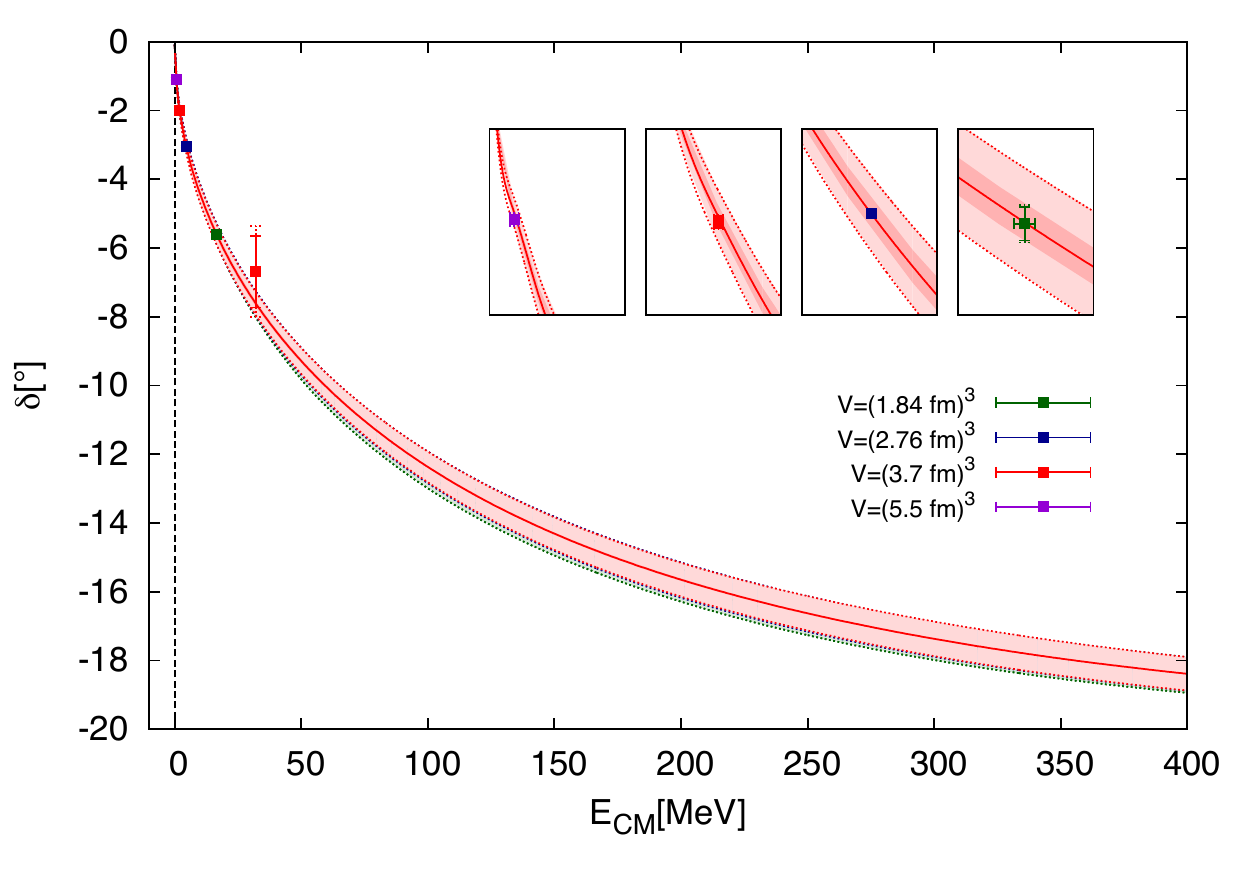}
\caption{
Potential in $I=2$ $\pi\pi$ channel (left panel) 
and $\delta(p)$ obtained from the potential (right panel) 
in quenched QCD calculated by Kurth {\it et al.}~\cite{Kurth:2013tua}.
}
\label{fig:LtoH:I2pipi}
\end{figure}

Kurth {\it et al.}~\cite{Kurth:2013tua} calculate $\delta(p)$
for the $I=2$ two-pion channel  by the two methods for $N_f = 0$ QCD at
$m_\pi = 940$ MeV on four volumes from $L=1.8$ to 5.5 fm.
The stability of $V(r)$ calculated by the $t$-dependent 
method of eq.(\ref{eq:tdep_v}) 
is illustrated in the left panel of Fig.~\ref{fig:LtoH:I2pipi}, where different colors correspond to different $t$ ranges, $t=15$--48($R_1$),
24--48($R_2$), and 33--48($R_3$).
The right panel of Fig.~\ref{fig:LtoH:I2pipi} compares $\delta(p)$ 
calculated from $V(r)$ (solid lines) with that extracted 
by L\"uscher's method (filled symbols). 
For this repulsive channel with a relatively simple shape of 
$V(r)$, we observe an agreement of the two methods for small momenta.

\subsection{$H$-dibaryon channel}

\begin{figure}[htbp]
\centering
\includegraphics[width=7cm,clip]{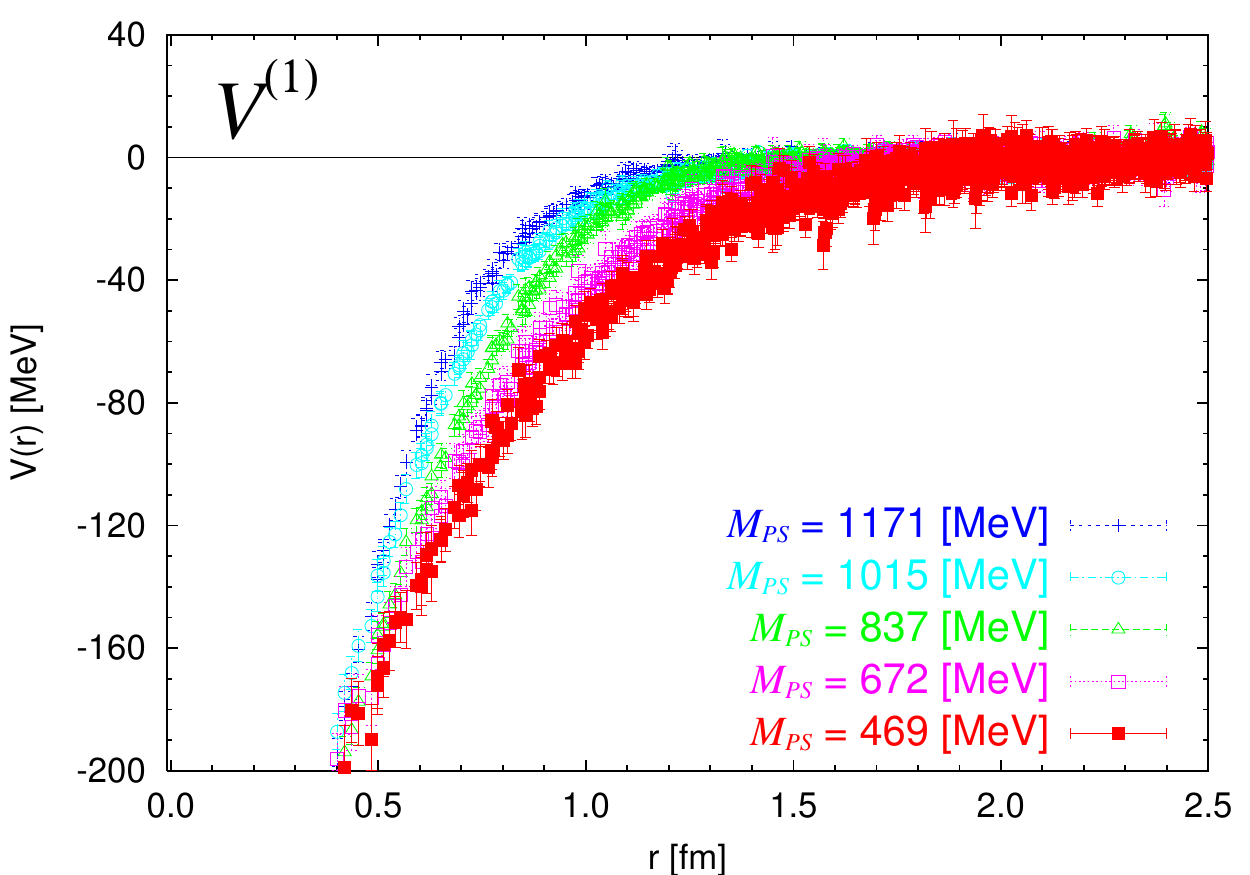}
\includegraphics[width=7.5cm,clip]{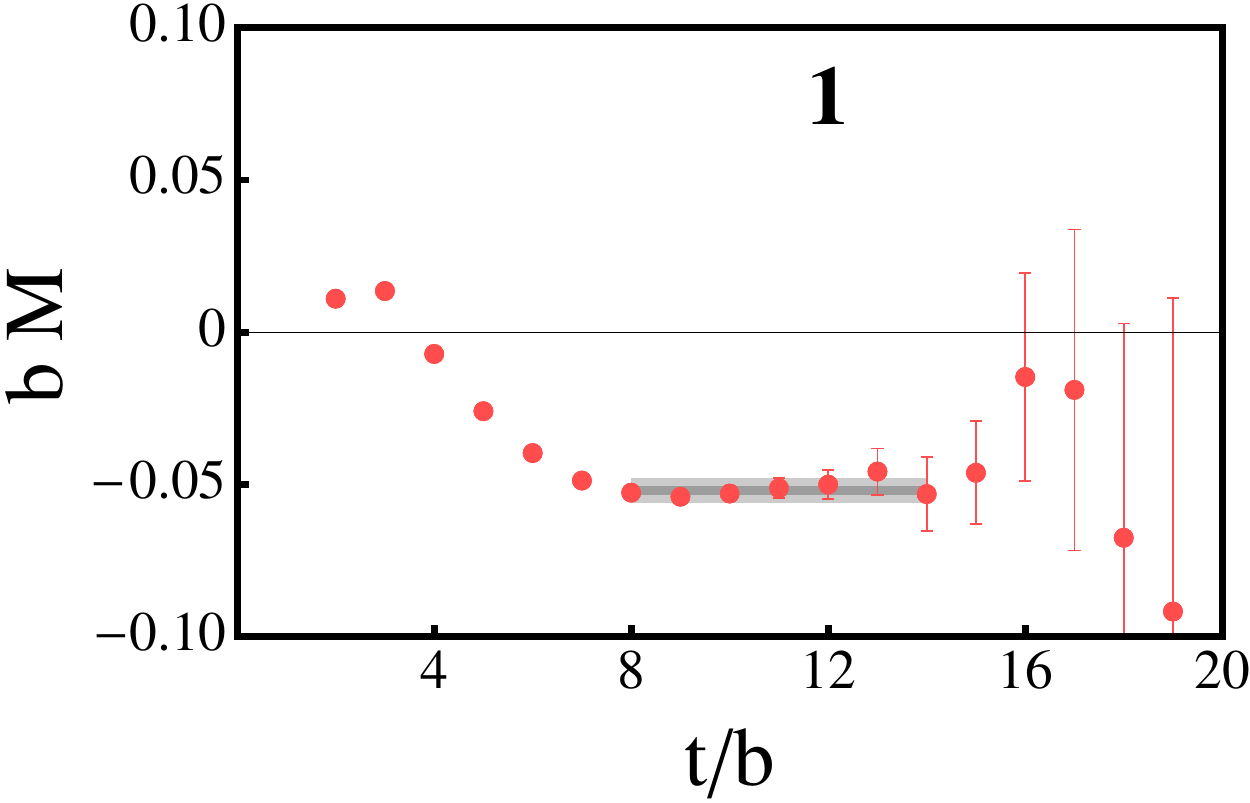}
\caption{
Potential calculated by 
HALQCD Collaboration~\cite{Inoue:2011ai}
(left panel) and effective energy shift calculated by
NPLQCD Collaboration~\cite{Beane:2012vq} (right panel)
in $N_f = 3$ H-dibaryon channel.
}
\label{fig:LtoH:Heach}
\end{figure}

HALQCD Collaboration~\cite{Inoue:2010es,Inoue:2011ai} examines the formation of 
H-dibaryon as a bound state in the $\Lambda\Lambda$ channel
in $N_f = 3$ QCD.
The result of $V(r)$ at $m_\pi$ in the range 470--1200~MeV obtained by the HALQCD method is shown in the left panel of Fig.~\ref{fig:LtoH:Heach}, 
which exhibits an attractive core in contrast to the $NN$ case.
Fitting $V(r)$ and solving the Schr\"odinger equation,
a bound state is found for all values of $m_\pi$ examined.

The right panel of Fig.~\ref{fig:LtoH:Heach} shows the result from
L\"uscher's method for $N_f = 3$ at $m_\pi = 800$ MeV
by NPLQCD Collaboration~\cite{Beane:2012vq}.
The effective energy shift of the ground $\Lambda\Lambda$ state from $2m_\Lambda$
is clearly non-zero in the plateau region.
After investigating the volume dependence of the energy shift,
it is concluded that the ground state is a bound state.

\begin{figure}[htbp]
\includegraphics[width=7cm,clip]{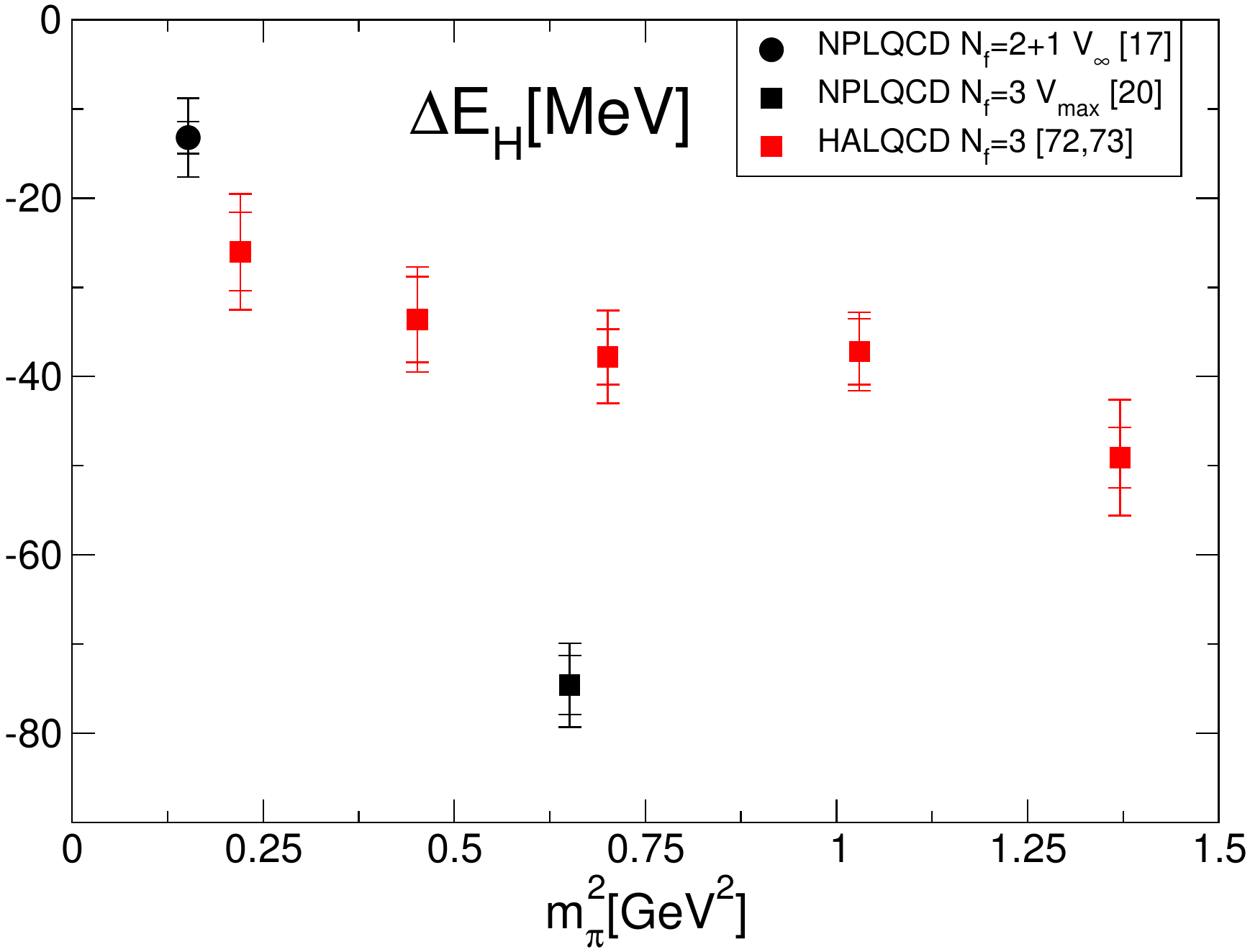}
\includegraphics[width=8.5cm,clip]{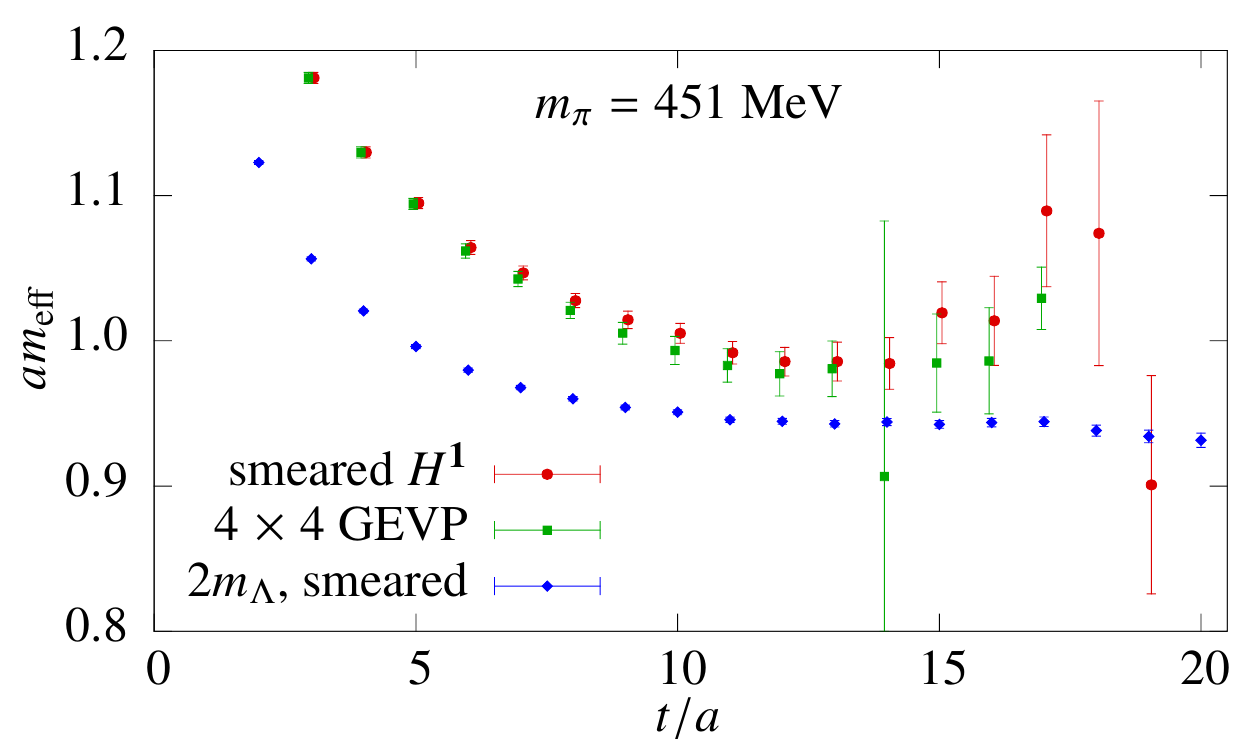}
\caption{
Binding energy obtained from L\"uscher's and HALQCD methods (left panel)
and preliminary result of effective energy shift calculated by
Green {\it et al.}~\cite{Green:2014dea} (right panel)
in H-dibaryon channel.
V$_{\infty}$ and V$_{\rm max}$ in the left panel express
results in the infinite volume and on the largest volume, respectively.
}
\label{fig:LtoH:H}
\end{figure}

The binding energies calculated in the two methods are plotted in the left
panel of Fig.~\ref{fig:LtoH:H} where we also added the result of 
NPLQCD Collaboration for $N_f = 2+1$~\cite{Beane:2010hg}. 
Qualitatively, the two methods are in agreement in that both predict bound states for the range of $m_\pi$ examined.  In quantitative detail, they differ significantly at $m_\pi \sim 800$ MeV.

In this conference, Green {\it et al.} report a preliminary
result~\cite{Green:2014dea} in $N_f = 2$ QCD
at $m_\pi = 1$ GeV and 450 MeV.
They employ the variational method~\cite{Luscher:1990ck} with
local products of six quark operator for both source and sink
time slices, and
conclude that the ground state has higher energy than
$2 m_\Lambda$, as presented in the right panel of Fig.~\ref{fig:LtoH:H} 
for the case of $m_\pi = 450$ MeV, in contrast to NPLQCD result.
A difference between NPLQCD's and their calculations is sink operator.
NPLQCD Collaboration uses two-baryon sink operator.
After the conference, a preliminary result of 
calculation with a two-baryon operator
is presented in other conference~\cite{Francis:2015bnl}, 
and they observe smaller ground state energy than $2m_\Lambda$.

\subsection{Two-nucleon channels}

Two nucleons can be in either in the spin-triplet $^3$S$_1$ channel or in the 
spin-singlet $^1$S$_0$; in nature the former has a deuteron,
while there are no bound states in the latter. 

\begin{figure}[htbp]
\centering
\includegraphics[width=7cm,clip]{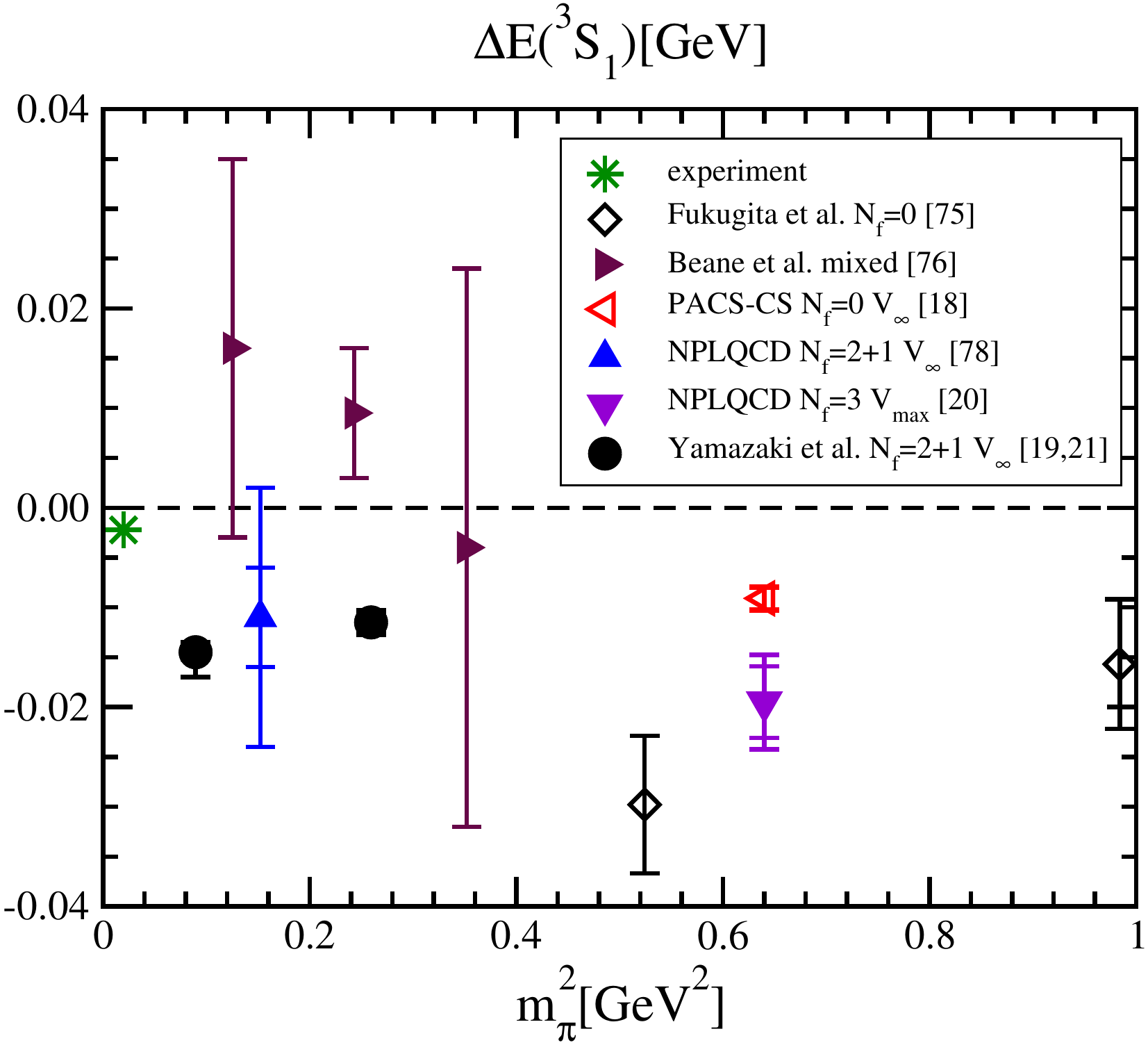}
\includegraphics[width=7cm,clip]{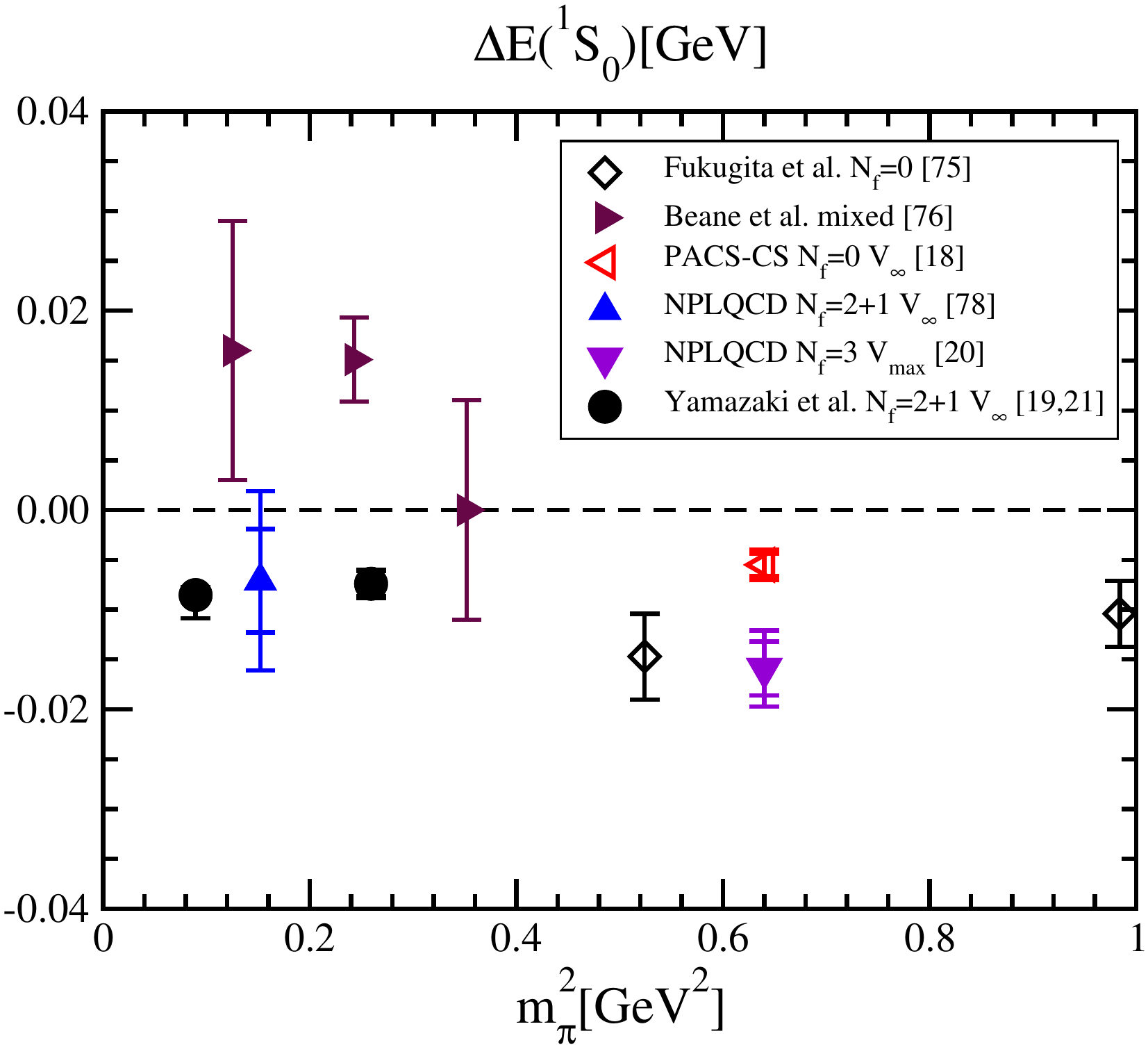}
\caption{
Energy shift in $NN$ $^3$S$_1$ (left) and $^1$S$_0$ (right) channels.
Experimental value of deuteron is also plotted in left panel.
V$_{\infty}$ and V$_{\rm max}$ express
results in the infinite volume and on the largest volume, respectively.
}
\label{fig:LtoH:NNdE}
\end{figure}

Figure~\ref{fig:LtoH:NNdE} summarizes all results
obtained so far by L\"uscher's method~\cite{Fukugita:1994ve,Beane:2006mx,Yamazaki:2011nd,Beane:2011iw,Beane:2012vq,Yamazaki:2012hi,Yamazaki:2015asa}
for the energy shift 
of the ground $NN$ state from $2m_N$
in the spin-triplet $^3$S$_1$ (left panel)
and spin-singlet $^1$S$_0$ (right panel) channels.
Early studies~\cite{Fukugita:1994ve,Beane:2006mx} did not examine
the volume dependence of the energy shift, 
so that it is not clear if the ground state is a scattering
state or a bound state.
Recent studies consciously examine this issue by working with multiple volumes~\cite{Yamazaki:2011nd,Beane:2011iw,Beane:2012vq,Yamazaki:2012hi,Yamazaki:2015asa}.
Except for one result~\cite{Beane:2011iw} (up triangle)
which unfortunately suffers from large error, 
all results obtained by study of volume dependence have found 
that the ground states in both the channels are bound states. 
The existence of those bound states at the large $m_\pi$ is also confirmed by
the sign of scattering length $a_0$ in each channel~\cite{Yamazaki:2011nd,Beane:2012vq}
based on the discussion in Refs.~\cite{Beane:2003da,Sasaki:2006jn}.

In the spin-triplet $^3$S$_1$ channel, the bound state nature found in lattice calculations is consistent with the presence of deuteron in nature.  The magnitude of binding energy, however, is significantly larger, and does not show a trend of decrease toward the experimental value as $m_\pi$ is taken lighter, at least to $m_\pi = 300$ MeV~\cite{Yamazaki:2015asa}.   
The trend is similar in the spin-singlet $^1$S$_0$ channel.  There is a bound state down to $m_\pi = 300$ MeV, and it is not clear if or how this bound state disappears as one approaches the physical pion mass.

A recent result for $\delta(p)$
in the $^1$S$_0$ channel obtained by the HALQCD method~\cite{Ishii:2013ira}
is shown in the right panel of Fig~\ref{fig:nf2+1:nforce}.
If there is a bound state, $\delta(p)$ should start at 180 degrees at $p=0$.
The figure therefore shows that there is no bound state
in this channel at $m_\pi = 410$--700 MeV.
In contrast to the results from L\"uscher's method described above,
this is consistent with experiment.  Quantitatively, however, the magnitude of phase shift in experiment is much larger than the values in Fig.~\ref{fig:nf2+1:nforce}.  It would be important to perform a calculation
at the physical $m_\pi$ to reproduce the experimental result.
Bound state is also not observed in the $^3$S$_1$ channel 
by the HALQCD method~\cite{Inoue:2011ai}.  
The $\delta(p)$, though increasing with decreasing $m_\pi$, 
does not support a bound state 
down to $m_\pi \approx 470$~MeV in $N_f=3$ QCD. 
This situation is opposite to that with the L\"uscher's method.

We have to conclude that at present the results from 
the two methods in the $NN$ channels
do not agree, even for the presence or absence of bound states.

\subsection{Systematic uncertainty}

We have seen that the results from the two methods 
agree with each other at least qualitatively
for the $I=2$ two-pion and $H$-dibaryon channels, while they differ in an essential way for the $NN$ channels. 
A possible reason for this difference may be that the first two cases are simple two-body systems 
governed by potentials which are monotone functions of $r$: 
$V(r)$ in the $I=2$ two-pion channel decreases as $r$ increases 
(Fig.~\ref{fig:LtoH:I2pipi}),
and the one in the $H$-dibaryon channel increases as $r$ (Fig.~\ref{fig:LtoH:H}).
The $NN$ potential, however, has a more complex structure (Fig.~\ref{fig:nf2+1:nforce}) with a repulsive core near the origin and an attractive tail at large distances.  
Thus, various systematic uncertainties might affect a balance toward bound state formation in a serious way. 
Let us then discuss possible systematic uncertainties in the two methods for the $NN$ channels.

\subsubsection{L\"uscher's method}

\begin{figure}[htbp]
\centering
\includegraphics[width=7cm,clip]{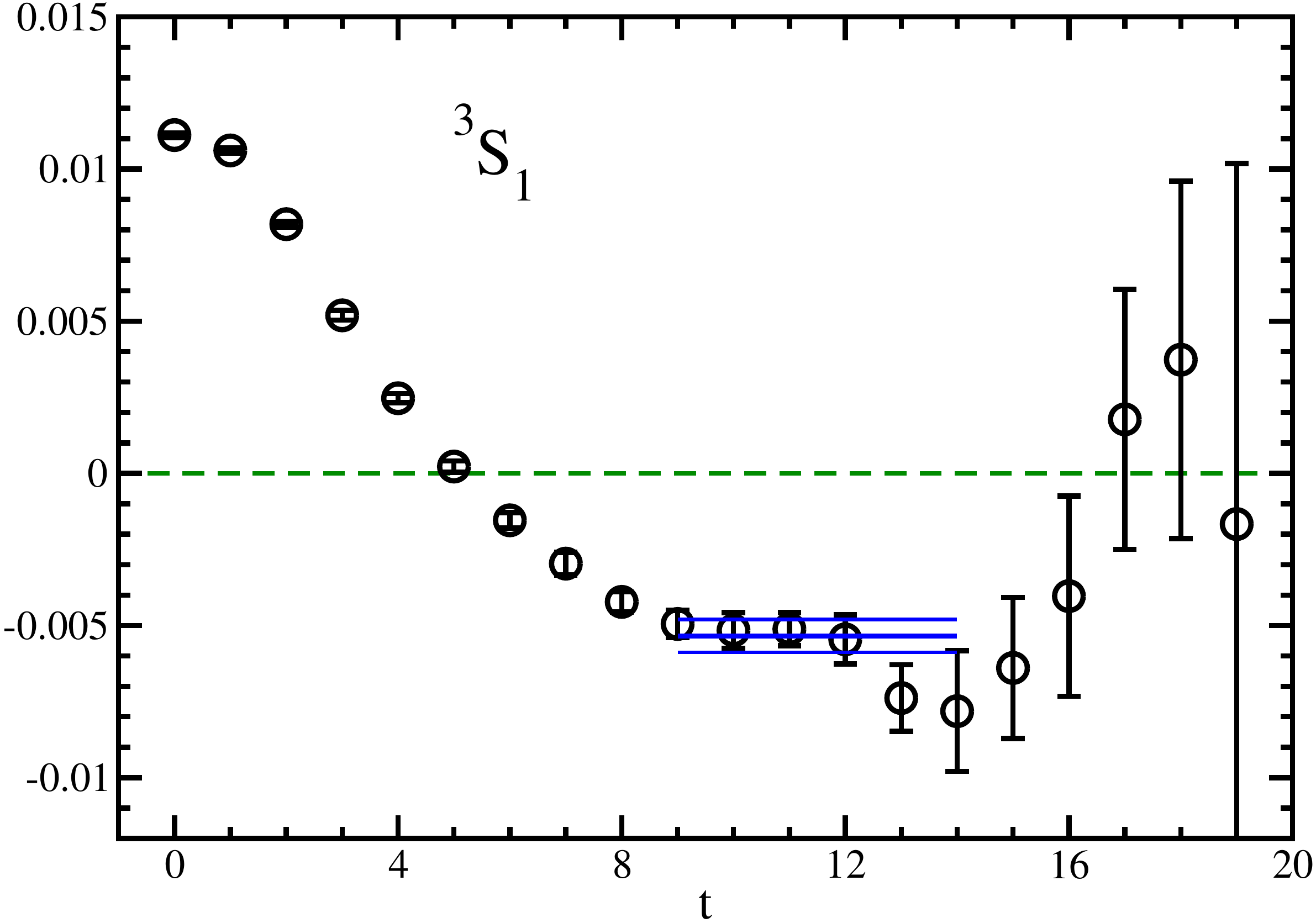}
\includegraphics[width=7.5cm,clip]{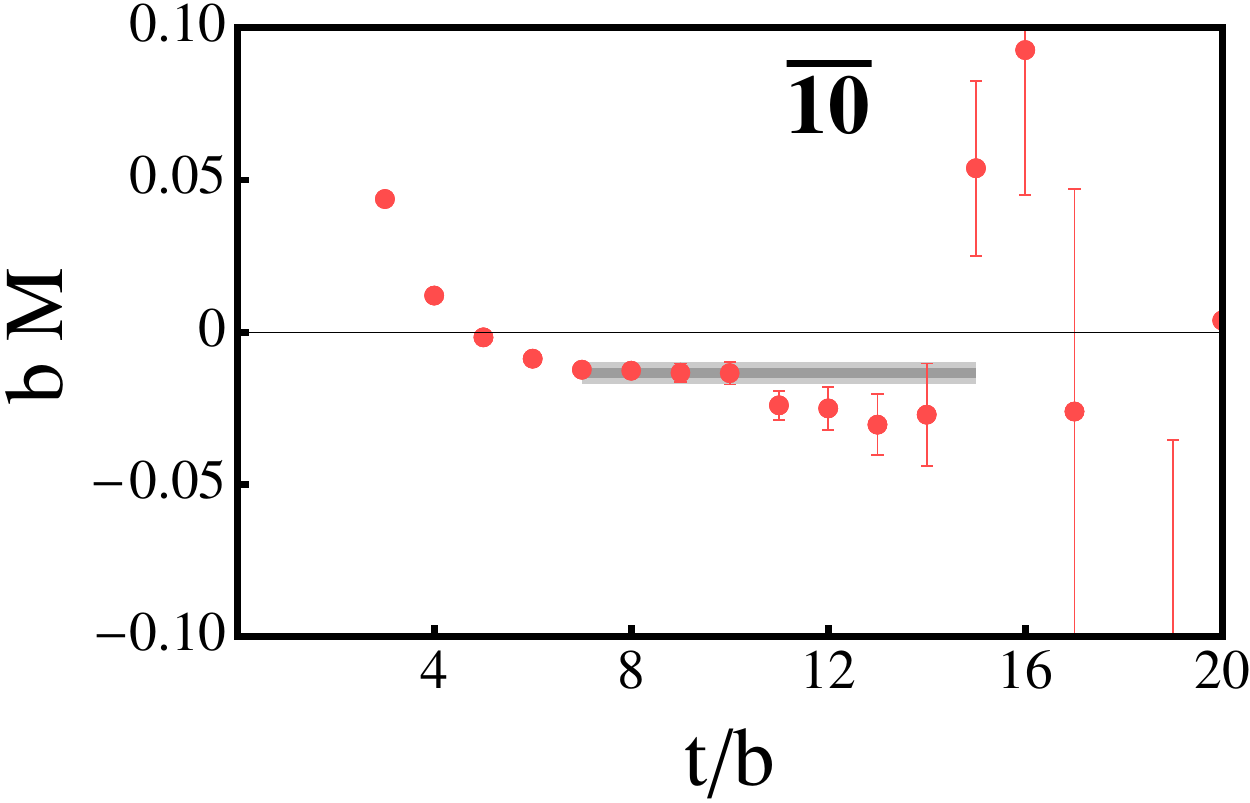}
\caption{
Effective energy shift calculated by 
Yamazaki {\it et al.}~\cite{Yamazaki:2012hi} (left)
and NPLQCD Collaboration~\cite{Beane:2012vq} (right)
in $NN$ $^3$S$_1$ channel.
}
\label{fig:LtoH:NNeff}
\end{figure}

For bound states, L\"uscher's method is nothing but the 
traditional calculation of evaluating the mass of the ground state 
of hadrons from two-point correlation function,
and extrapolating it to the infinite volume.
Thus, the crucial point we need to worry about is how reliably 
the ground state energy is extracted for a given volume.
Figure~\ref{fig:LtoH:NNeff} presents the
effective energy shift for the $NN$ state from $2m_N$ in
the $^3$S$_1$ channel
by Yamazaki {\it et al.}~\cite{Yamazaki:2012hi}
and NPLQCD Collaboration~\cite{Beane:2012vq}.
While a clear plateau extends only over 3 or 4 time slices, 
the negative value of the energy shift appears quite convincing. 

One should note that the results above are obtained with only one type of source operator.  It is a legitimate question to ask if one obtains the same result, when a different source operator is employed.  Similarly, the dependence on the choice of the sink operator is also a potential issue.  These issues have not been thoroughly investigated.  Application of the variational method~\cite{Luscher:1990ck} would be worthwhile to settle these issues. 

There are two more sources of possible errors which, however, are more dynamical in nature.  One is the dependence on the quark mass.  It may well be that the bound state found in the $^1$S$_0$ channel for the pion mass down to $m_\pi \approx 300$ MeV~\cite{Yamazaki:2015asa} disappears at a smaller pion mass but heavier than the physical mass.  

Another is the dependence on the lattice spacing.  It is possible in principle that the pion mass $m_\pi^c$ below which a bound state in the spin-singlet $^1$S$_0$ channel disappears varies with the lattice spacing.  It may require a sufficiently small lattice spacing before $m_\pi^c$ becomes non-zero or larger than the physical value.  

These possibilities can only be settled by repeating the calculations toward the physical pion mass and for smaller lattice spacings.

\subsubsection{HALQCD method}

The HALQCD method relies crucially on the potential $V(\vec{r})$ extracted from the NBS wave function.  Hence possible systematic errors center around the meaning of the potential and how the final physics results may or may not depend on the ambiguities in the potential. 
Let us discuss possible sources of systematic errors and ambiguities one by one. 

\noindent  (i) Sink operator dependence of $V(\vec{r})$ and $\delta(p)$\\
The potential $V(\vec{r})$ will in general depend on the choice of 
the sink operator of $C_{NN}(\vec{r},t)$ in eq.(\ref{eq:fourpt}). 
In SU(2) gauge theory, Takahashi {\it et al.}~\cite{Takahashi:2009ef} examined this issue, and found that 
$V(\vec{r})$ in the small $r$ region largely depends on the choice of 
the sink operator, while the dependence is small for large $r$. 
To what extent such a dependence affects the phase shift extracted from $V(\vec{r})$ was not tested in this study, 
but this is certainly a point which requires a detailed check. 

\noindent (ii) Velocity expansion of $U(\vec{r},\vec{r}^\prime)$\\
If higher order of velocity expansion is not negligible in
eq.(\ref{eq:assump_v}), $V(\vec{r})$ obtained from eq.(\ref{eq:nlp})
or (\ref{eq:tdep_v})
depends on the relative momentum $p$ of the two-nucleon state.
The dependency was investigated by Murano {\it et al.}~\cite{Murano:2011nz}
at $p \sim 0$ and 250 MeV, and it was concluded that
the $V(\vec{r})$ obtained from two values of $p$ agree
within the statistical error.

\noindent (iii)  Volume dependence of $V(\vec{r})$\\
If the extracted potential $V(\vec{r})$ has finite volume effects, it may affect $\delta(p)$ and hence the presence or absence of bound states. 
Thus, finite volume dependence was investigated by
HALQCD Collaboration~\cite{Inoue:2010es} at $m_\pi = 1$ GeV
in the spatial extent of 2--3 fm,
and it was concluded that the finite volume effects in $V(\vec{r})$ is
negligible within the statistical error.

\noindent(iv)  Lattice spacing dependence of $V(\vec{r})$\\
Lattice spacing dependence of $V(\vec{r})$ was examined at
$m_\pi = 1.1$ GeV for three different lattice spacings, $a=0.11$--0.22 fm,
by HALQCD Collaboration~\cite{Doi:2013haa}.
They found that lattice spacing dependence is small in the larger $r$ region,
but it is large in the small $r$ region, $r \simlt 0.5$ fm,
and then $\delta(p)$'s obtained at different lattice spacings
differ at large $p$ region.

\noindent (v) Determination of $m_N$\\
If the value of $m_N$ contains
systematic errors, which is even a few MeV,
it corresponds to a constant shift of $V(r)$ for all values of $r$, since 
\begin{equation}
V({\vec r}) \approx \frac{\left(\frac{\nabla^2}{m_N}
- \frac{\partial}{\partial t}\right)
C_{NN}({\vec r}, t)}{C_{NN}({\vec r}, t)} - 2m_N .
\end{equation}
The tiny constant shift of $V(r)$ might be crucial in some case, {\it e.g.},
identification of existence of bound state.

\noindent (vi) $V(\vec{r})$ in large $r$ region\\
It is hard to extract $V(\vec{r})$ in the large $r$ region precisely,
because the noise-to-signal ratio of $V(\vec{r})$
increases with $r$.
Since the large $r$ region is important to determine
how long attractive tail of the potential continues, this uncertainty
might affect existence of bound state as pointed out by
the plenary speaker in the last conference~\cite{Walker-Loud:2014iea}.

To sum up the considerations above, most uncertain with the HALQCD method seems to be the ambiguities in the potential for small values of $r$ as written in (i), (iii), and (iv).  
The small $r$ region is not relevant to calculate $\delta(p)$
at small $p$, while it becomes important to calculate $\delta(p)$
at large $p$, and also to test the existence of a bound state.
Thus, those uncertainties in the small $r$ region might cause
the difference of the result obtained from the two methods.

\section{Light nuclei}
\label{sec:nuclei}

A first attempt to simulate the formation of helium nuclei in lattice QCD 
was  reported in Ref.~\cite{Yamazaki:2009ua}.
In this work, the binding energies of the $^4$He and $^3$He nuclei were
calculated in quenched QCD at $m_\pi=800$ MeV, 
by examining the volume dependence of the energy shift. 
A serious computational problem with nuclei calculations is 
huge number of Wick contraction.
For helium nuclei~\cite{Yamazaki:2009ua},
this problem was overcome 
by omitting calculations of redundant contractions under symmetries
of interpolating operator,
and by utilizing blocks of three quark propagators.
Recently, more efficient calculation 
methods~\cite{Doi:2012xd,Detmold:2012eu,Gunther:2013xj} have been proposed.
In this conference a preliminary result in 
this direction~\cite{Vachaspati:2014bda}
is reported.

\begin{figure}[htbp]
\centering
\includegraphics[width=7cm,clip]{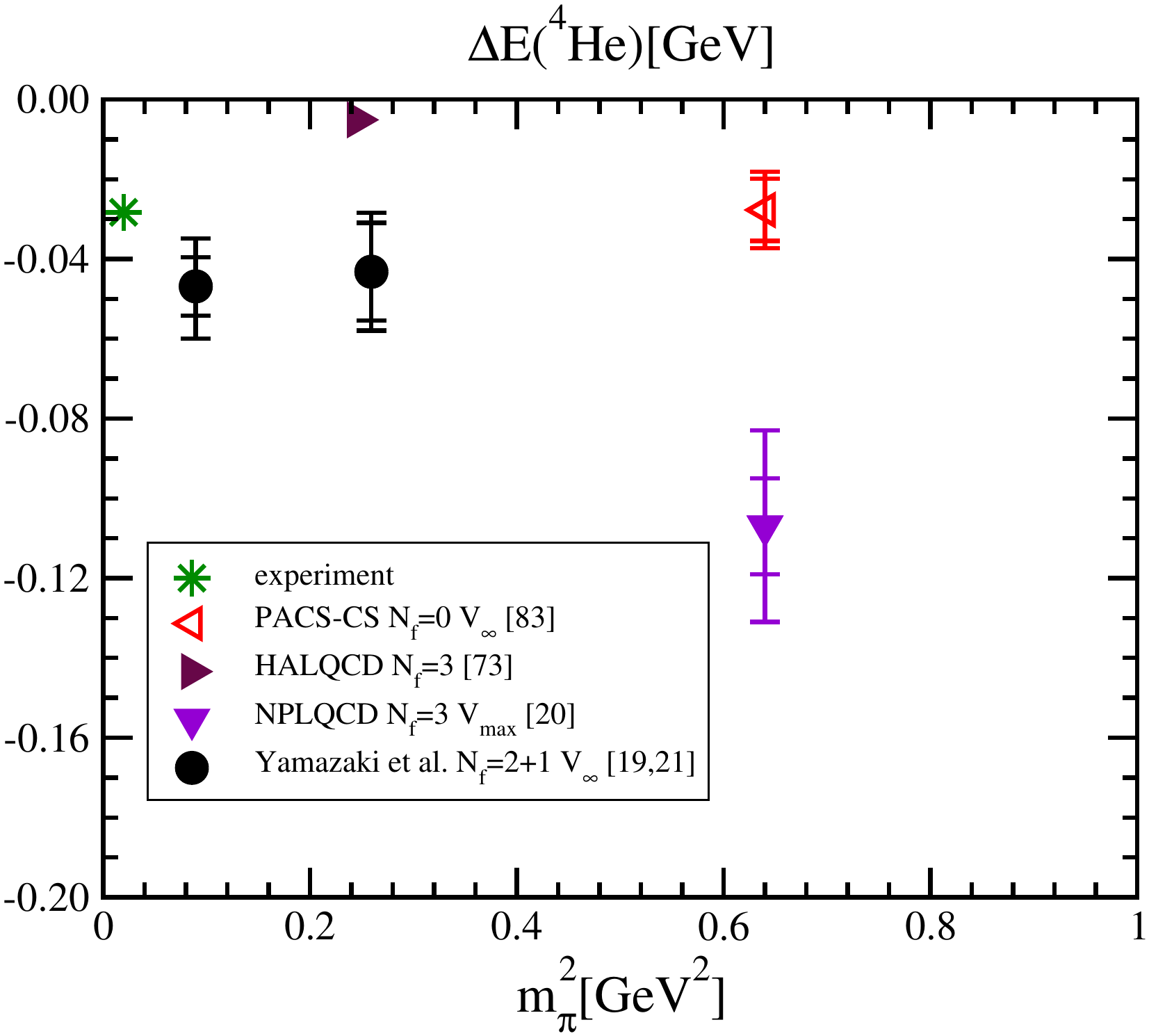}
\includegraphics[width=7cm,clip]{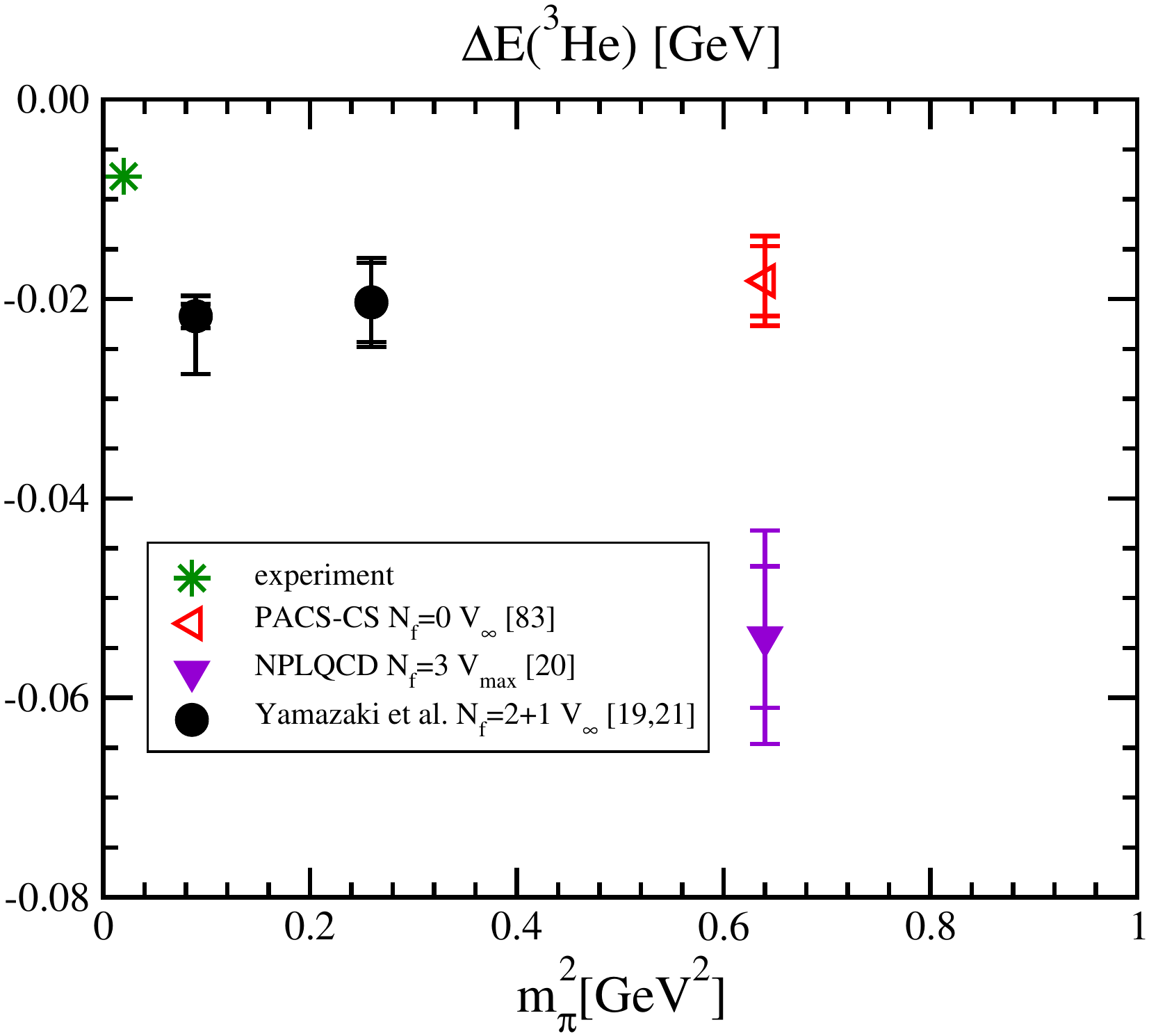}
\caption{
Binding energies for $^4$He (left) and $^3$He (right) channels.
Experimental values are also plotted.
V$_{\infty}$ and V$_{\rm max}$ express
results in the infinite volume and on the largest volume, respectively.
}
\label{fig:nuclei}
\end{figure}

The exploratory study of helium nuclei 
has been followed by calculations in
$N_f = 3$ QCD at $m_\pi = 810$ MeV~\cite{Beane:2012vq},
and $2+1$ QCD at $m_\pi = 510$ MeV~\cite{Yamazaki:2012hi} and
$m_\pi = 300$ MeV~\cite{Yamazaki:2015asa}.
Results for the binding energies for $^4$He and $^3$He nuclei are summarized
in Fig.~\ref{fig:nuclei}.
For both the nuclei, the quenched and $N_f = 3$ calculations
at $m_\pi \sim 800$ MeV give different results.
While it might be a dynamical quark effect,
precise understanding is lacking at present. 
At $m_\pi = 300$, the lightest pion explored so far, the binding energy of $^4$He
is roughly consistent with the experiment, while
that for $^3$He is about three times larger than the experiment.
The discrepancy between the lattice and experimental results might
be caused by a large $m_\pi$ in the calculation.
Future calculations closer to the physical $m_\pi$ should tell if this expectation is justified. 

In the right panel of Fig.~\ref{fig:nuclei} the result from HALQCD method for 
$^4$He in $N_f = 3$ QCD~\cite{Inoue:2011ai}, 
giving a very small binding energy, is also plotted. 
Results for $^3$He is not available from the HALQCD method.

Let us list some related studies for nuclei. 
NPLQCD Collaboration calculates 
magnetic moments for light nuclei~\cite{Beane:2014ora}
and binding energies for
quarkonium-nucleus bound states~\cite{Beane:2014sda}.
HALQCD Collaboration calculates the binding energy of 
spin-2 $N\Omega$ bound state~\cite{Etminan:2014tya}
and medium-heavy nuclei~\cite{Inoue:2011ai}
from potentials obtained by the HALQCD method.
The nuclei calculation has been extended to theories beyond the standard model.
Detmold {\it et al.}~\cite{Detmold:2014kba} 
calculate nuclei of two to four particles in two-flavor
SU(2) gauge theory to explore the importance of nuclear physics 
in the strongly-interacting dark matter models.

\section{Conclusion}
\label{sec:conclusion}

In this report we review recent works for scattering and decay channels.
The scattering length in $I=2$ two-pion channel is
precisely determined using chiral extrapolation with ChPT formula.
Still careful estimation of systematic errors are necessary,
some recent results have smaller error than the experimental results.
Thus, the calculation in this channel enters precision measurement
era as in hadron masses and decay constants.
The calculation of scattering length in $I=1/2$ $K\pi$ channel
is more difficult than the one in $I=2$ two-pion channel,
so that only a few calculations are carried out.
The resonance mass and decay width are extracted from
scattering phase shift.
In $I=1$ two-pion channel, the $\rho$ meson mass
and its coupling to two pions are calculated.
Both the physical quantities are roughly consistent with the experiment
within large error of lattice results.

Comparisons of the results from L\"uscher's method and HALQCD method
are discussed.
In $I=2$ two-pion channel, the results from the two methods
agree with each other numerically.
While the binding energies of $H$-dibaryon differ in larger $m_\pi$ region,
the two method give the consistent conclusion that
the $H$-dibaryon exists in larger $m_\pi$ than physical one.
However, in the $NN$ channels, the two methods obtain
opposite results in the point of view of the presence of bound states.
Possible uncertainties in each method are discussed in the $NN$ channels.
Contaminations from higher excited states,
larger quark mass, and finite lattice spacing could affect
the result from L\"uscher's method.
For HALQCD method, several possible uncertainties are listed.
Uncertainties of the potential in small $r$ region
might explain why the bound state is not formed in this method.
In order to understand the difference between the results
obtained from L\"uscher's method and HALQCD method,
it would be necessary to carry out more detail investigations for
uncertainties in both the methods. 
Furthermore, understanding of relation between
the potential evaluated from NBS wave function,
in other words, Fourier transformation of off-shell scattering amplitude, 
and phase shift in quantum field theory might shed light on 
the difference.

The helium nuclei are calculated in several works using L\"uscher's method.
In the $^4$He channel the binding energy is roughly consistent with
the experiment, while in the $^3$He channel it is larger than the
experiment.
The discrepancy of the binding energy from the experiment is
considered to be mainly caused by large $m_\pi$ in the calculation.
Therefore, calculations at smaller $m_\pi < 300$ MeV would be
necessary to study whether lattice calculation reproduces
the experimental binding energy or not.

\section*{Acknowledgements}
I would like to thank
John Bulava, Raul Brice\~no, William Detmold, Jeremy Green, 
Yoshinobu Kuramashi, 
Christian Lang, Thibaut Metivet, Colin Morningstar,
Hidekatsu Nemura, Sasa Prelovsek, and Akira Ukawa
for discussions and sending comments and figures for this talk.
This work is partially supported by
the JSPS Grant-in-Aid for Scientific Research 
for Young Scientists (B) No.25800138, and
also by Grants-in-Aid of the Japanese Ministry for 
Scientific Research on Innovative Areas No.23105708.

\bibliography{lat14}

\providecommand{\href}[2]{#2}\begingroup\raggedright\begin{thebibliography}{10}

\bibitem{Luscher:1986pf}
M.~L{\"u}scher, {\em Commun. Math. Phys.} {\bf 105} (1986) 153--188.

\bibitem{Luscher:1990ux}
M.~L{\"u}scher, {\em Nucl. Phys.} {\bf B354} (1991) 531--578.

\bibitem{Rummukainen:1995vs}
K.~Rummukainen and S.~A. Gottlieb, {\em Nucl.Phys.} {\bf B450} (1995) 397--436,
  [\href{http://xxx.lanl.gov/abs/hep-lat/9503028}{{\tt hep-lat/9503028}}].

\bibitem{Kim:2005gf}
C.~Kim, C.~Sachrajda, and S.~R. Sharpe, {\em Nucl.Phys.} {\bf B727} (2005)
  218--243, [\href{http://xxx.lanl.gov/abs/hep-lat/0507006}{{\tt
  hep-lat/0507006}}].

\bibitem{Christ:2005gi}
N.~H. Christ, C.~Kim, and T.~Yamazaki, {\em Phys.Rev.} {\bf D72} (2005) 114506,
  [\href{http://xxx.lanl.gov/abs/hep-lat/0507009}{{\tt hep-lat/0507009}}].

\bibitem{Feng:2011ah}
{\bf ETM Collaboration}, X.~Feng, K.~Jansen, and D.~B. Renner, {\em PoS} {\bf
  LATTICE2010} (2010) 104, [\href{http://xxx.lanl.gov/abs/1104.0058}{{\tt
  arXiv:1104.0058}}].

\bibitem{Dudek:2012gj}
J.~J. Dudek, R.~G. Edwards, and C.~E. Thomas, {\em Phys.Rev.} {\bf D86} (2012)
  034031, [\href{http://xxx.lanl.gov/abs/1203.6041}{{\tt arXiv:1203.6041}}].

\bibitem{Fu:2011xz}
Z.~Fu, {\em Phys.Rev.} {\bf D85} (2012) 014506,
  [\href{http://xxx.lanl.gov/abs/1110.0319}{{\tt arXiv:1110.0319}}].

\bibitem{Leskovec:2012gb}
L.~Leskovec and S.~Prelovsek, {\em Phys.Rev.} {\bf D85} (2012) 114507,
  [\href{http://xxx.lanl.gov/abs/1202.2145}{{\tt arXiv:1202.2145}}].

\bibitem{Doring:2012eu}
M.~Doring, U.~Meissner, E.~Oset, and A.~Rusetsky, {\em Eur.Phys.J.} {\bf A48}
  (2012) 114, [\href{http://xxx.lanl.gov/abs/1205.4838}{{\tt
  arXiv:1205.4838}}].

\bibitem{Gockeler:2012yj}
M.~Gockeler, R.~Horsley, M.~Lage, U.-G. Meissner, P.~Rakow, {\em et.~al.}, {\em
  Phys.Rev.} {\bf D86} (2012) 094513,
  [\href{http://xxx.lanl.gov/abs/1206.4141}{{\tt arXiv:1206.4141}}].

\bibitem{Li:2012bi}
N.~Li and C.~Liu, {\em Phys.Rev.} {\bf D87} (2013) 014502,
  [\href{http://xxx.lanl.gov/abs/1209.2201}{{\tt arXiv:1209.2201}}].

\bibitem{Briceno:2014pka}
R.~A. Brice\~no, \href{http://xxx.lanl.gov/abs/1411.6944}{{\tt
  arXiv:1411.6944}}.

\bibitem{Prelovsek:2014zga}
S.~Prelovsek, {\em PoS} {\bf LATTICE2014} (2014) 015,
  [\href{http://xxx.lanl.gov/abs/1411.0405}{{\tt arXiv:1411.0405}}].

\bibitem{Beane:2003da}
S.~R. Beane, P.~F. Bedaque, A.~Parreno, and M.~J. Savage, {\em Phys. Lett.}
  {\bf B585} (2004) 106--114,
  [\href{http://xxx.lanl.gov/abs/hep-lat/0312004}{{\tt hep-lat/0312004}}].

\bibitem{Sasaki:2006jn}
S.~Sasaki and T.~Yamazaki, {\em Phys. Rev.} {\bf D74} (2006) 114507,
  [\href{http://xxx.lanl.gov/abs/hep-lat/0610081}{{\tt hep-lat/0610081}}].

\bibitem{Beane:2010hg}
{\bf NPLQCD Collaboration}, S.~Beane {\em et.~al.}, {\em Phys.Rev.Lett.} {\bf
  106} (2011) 162001, [\href{http://xxx.lanl.gov/abs/1012.3812}{{\tt
  arXiv:1012.3812}}].

\bibitem{Yamazaki:2011nd}
{\bf PACS-CS Collaboration}, T.~Yamazaki, Y.~Kuramashi, and A.~Ukawa, {\em
  Phys. Rev.} {\bf D84} (2011) 054506,
  [\href{http://xxx.lanl.gov/abs/1105.1418}{{\tt arXiv:1105.1418}}].

\bibitem{Yamazaki:2012hi}
T.~Yamazaki, K.-i. Ishikawa, Y.~Kuramashi, and A.~Ukawa, {\em Phys.Rev.} {\bf
  D86} (2012) 074514, [\href{http://xxx.lanl.gov/abs/1207.4277}{{\tt
  arXiv:1207.4277}}].

\bibitem{Beane:2012vq}
S.~Beane, E.~Chang, S.~Cohen, W.~Detmold, H.~Lin, {\em et.~al.}, {\em
  Phys.Rev.} {\bf D87} (2013), no.~3 034506,
  [\href{http://xxx.lanl.gov/abs/1206.5219}{{\tt arXiv:1206.5219}}].

\bibitem{Yamazaki:2015asa}
T.~Yamazaki, K.-i. Ishikawa, Y.~Kuramashi, and A.~Ukawa,
  \href{http://xxx.lanl.gov/abs/1502.0418}{{\tt arXiv:1502.0418}}.

\bibitem{Yamazaki:2004qb}
{\bf CP-PACS Collaboration}, T.~Yamazaki {\em et.~al.}, {\em Phys. Rev.} {\bf
  D70} (2004) 074513, [\href{http://xxx.lanl.gov/abs/hep-lat/0402025}{{\tt
  hep-lat/0402025}}].

\bibitem{Beane:2005rj}
{\bf NPLQCD Collaboration}, S.~R. Beane, P.~F. Bedaque, K.~Orginos, and M.~J.
  Savage, {\em Phys.Rev.} {\bf D73} (2006) 054503,
  [\href{http://xxx.lanl.gov/abs/hep-lat/0506013}{{\tt hep-lat/0506013}}].

\bibitem{Beane:2007xs}
{\bf NPLQCD Collaboration}, S.~R. Beane, T.~C. Luu, K.~Orginos, A.~Parreno,
  M.~J. Savage, {\em et.~al.}, {\em Phys.Rev.} {\bf D77} (2008) 014505,
  [\href{http://xxx.lanl.gov/abs/0706.3026}{{\tt arXiv:0706.3026}}].

\bibitem{Liu:2009uw}
{\bf RBC-UKQCD Collaboration}, Q.~Liu, {\em PoS} {\bf LAT2009} (2009) 101,
  [\href{http://xxx.lanl.gov/abs/0910.2658}{{\tt arXiv:0910.2658}}].

\bibitem{Feng:2009ij}
X.~Feng, K.~Jansen, and D.~B. Renner, {\em Phys.Lett.} {\bf B684} (2010)
  268--274, [\href{http://xxx.lanl.gov/abs/0909.3255}{{\tt arXiv:0909.3255}}].

\bibitem{Yagi:2011jn}
T.~Yagi, S.~Hashimoto, O.~Morimatsu, and M.~Ohtani,
  \href{http://xxx.lanl.gov/abs/1108.2970}{{\tt arXiv:1108.2970}}.

\bibitem{Fu:2011bz}
Z.~Fu, {\em Commun.Theor.Phys.} {\bf 57} (2012) 78--84,
  [\href{http://xxx.lanl.gov/abs/1110.3918}{{\tt arXiv:1110.3918}}].

\bibitem{Fu:2013ffa}
Z.~Fu, {\em Phys.Rev.} {\bf D87} (2013) 074501,
  [\href{http://xxx.lanl.gov/abs/1303.0517}{{\tt arXiv:1303.0517}}].

\bibitem{Sasaki:2013vxa}
{\bf PACS-CS Collaboration}, K.~Sasaki, N.~Ishizuka, M.~Oka, and T.~Yamazaki,
  {\em Phys.Rev.} {\bf D89} (2014) 054502,
  [\href{http://xxx.lanl.gov/abs/1311.7226}{{\tt arXiv:1311.7226}}].

\bibitem{Pislak:2003sv}
S.~Pislak, R.~Appel, G.~Atoyan, B.~Bassalleck, D.~Bergman, {\em et.~al.}, {\em
  Phys.Rev.} {\bf D67} (2003) 072004,
  [\href{http://xxx.lanl.gov/abs/hep-ex/0301040}{{\tt hep-ex/0301040}}].
  [Erratum-ibid.\ {\bf D81} (2010) 119903].

\bibitem{Batley:2010zza}
{\bf NA48-2 Collaboration}, J.~Batley {\em et.~al.}, {\em Eur.Phys.J.} {\bf
  C70} (2010) 635--657.

\bibitem{Dudek:2010ew}
{\bf Hadron Spectrum Collaboration}, J.~J. Dudek, R.~G. Edwards, M.~J. Peardon,
  D.~G. Richards, and C.~E. Thomas, {\em Phys.Rev.} {\bf D83} (2011) 071504,
  [\href{http://xxx.lanl.gov/abs/1011.6352}{{\tt arXiv:1011.6352}}].

\bibitem{Beane:2011sc}
{\bf NPLQCD Collaboration}, S.~Beane {\em et.~al.}, {\em Phys.Rev.} {\bf D85}
  (2012) 034505, [\href{http://xxx.lanl.gov/abs/1107.5023}{{\tt
  arXiv:1107.5023}}].

\bibitem{Colangelo:2001df}
G.~Colangelo, J.~Gasser, and H.~Leutwyler, {\em Nucl.Phys.} {\bf B603} (2001)
  125--179, [\href{http://xxx.lanl.gov/abs/hep-ph/0103088}{{\tt
  hep-ph/0103088}}].

\bibitem{Helmes:2014wca}
C.~Helmes, C.~Jost, B.~Knippschild, L.~Liu, C.~Urbach, {\em et.~al.},
  \href{http://xxx.lanl.gov/abs/1412.0408}{{\tt arXiv:1412.0408}}.

\bibitem{Morningstar:2011ka}
C.~Morningstar, J.~Bulava, J.~Foley, K.~J. Juge, D.~Lenkner, {\em et.~al.},
  {\em Phys.Rev.} {\bf D83} (2011) 114505,
  [\href{http://xxx.lanl.gov/abs/1104.3870}{{\tt arXiv:1104.3870}}].

\bibitem{Aoki:2011yj}
{\bf PACS-CS Collaboration}, S.~Aoki {\em et.~al.}, {\em Phys.Rev.} {\bf D84}
  (2011) 094505, [\href{http://xxx.lanl.gov/abs/1106.5365}{{\tt
  arXiv:1106.5365}}].

\bibitem{Nagata:2008wk}
J.~Nagata, S.~Muroya, and A.~Nakamura, {\em Phys.Rev.} {\bf C80} (2009) 045203,
  [\href{http://xxx.lanl.gov/abs/0812.1753}{{\tt arXiv:0812.1753}}].

\bibitem{Fu:2011wc}
Z.~Fu, {\em Phys.Rev.} {\bf D85} (2012) 074501,
  [\href{http://xxx.lanl.gov/abs/1110.1422}{{\tt arXiv:1110.1422}}].

\bibitem{Lang:2012sv}
C.~Lang, L.~Leskovec, D.~Mohler, and S.~Prelovsek, {\em Phys.Rev.} {\bf D86}
  (2012) 054508, [\href{http://xxx.lanl.gov/abs/1207.3204}{{\tt
  arXiv:1207.3204}}].

\bibitem{Buettiker:2003pp}
P.~Buettiker, S.~Descotes-Genon, and B.~Moussallam, {\em Eur.Phys.J.} {\bf C33}
  (2004) 409--432, [\href{http://xxx.lanl.gov/abs/hep-ph/0310283}{{\tt
  hep-ph/0310283}}].

\bibitem{Beane:2006gj}
{\bf NPLQCD Collaboration}, S.~R. Beane, P.~F. Bedaque, T.~C. Luu, K.~Orginos,
  E.~Pallante, {\em et.~al.}, {\em Phys.Rev.} {\bf D74} (2006) 114503,
  [\href{http://xxx.lanl.gov/abs/hep-lat/0607036}{{\tt hep-lat/0607036}}].

\bibitem{Janowski:2014lat}
{\bf RBC-UKQCD Collaboration}, T.~Janowski.
\newblock
  https://indico.bnl.gov/getFile.py/access?contribId=130\&sessionId=2\&resId=0\&materialId=slides\&confId=736.

\bibitem{Wilson:2014cna}
{\bf Hadron Spectrum Collaboration}, D.~J. Wilson, J.~J. Dudek, R.~G. Edwards,
  and C.~E. Thomas, \href{http://xxx.lanl.gov/abs/1411.2004}{{\tt
  arXiv:1411.2004}}.

\bibitem{Luscher:1990ck}
M.~L{\"u}scher and U.~Wolff, {\em Nucl. Phys.} {\bf B339} (1990) 222--252.

\bibitem{Aoki:2007rd}
{\bf CP-PACS Collaboration}, S.~Aoki {\em et.~al.}, {\em Phys.Rev.} {\bf D76}
  (2007) 094506, [\href{http://xxx.lanl.gov/abs/0708.3705}{{\tt
  arXiv:0708.3705}}].

\bibitem{Feng:2010es}
{\bf ETM Collaboration}, X.~Feng, K.~Jansen, and D.~B. Renner, {\em Phys.Rev.}
  {\bf D83} (2011) 094505, [\href{http://xxx.lanl.gov/abs/1011.5288}{{\tt
  arXiv:1011.5288}}].

\bibitem{Lang:2011mn}
C.~Lang, D.~Mohler, S.~Prelovsek, and M.~Vidmar, {\em Phys.Rev.} {\bf D84}
  (2011), no.~5 054503, [\href{http://xxx.lanl.gov/abs/1105.5636}{{\tt
  arXiv:1105.5636}}]. [Erratum-ibid.\ {\bf D89} (2014), no. 5 059903].

\bibitem{Pelissier:2012pi}
C.~Pelissier and A.~Alexandru, {\em Phys.Rev.} {\bf D87} (2013) 014503,
  [\href{http://xxx.lanl.gov/abs/1211.0092}{{\tt arXiv:1211.0092}}].

\bibitem{Dudek:2012xn}
{\bf Hadron Spectrum Collaboration}, J.~J. Dudek, R.~G. Edwards, and C.~E.
  Thomas, {\em Phys.Rev.} {\bf D87} (2013), no.~3 034505,
  [\href{http://xxx.lanl.gov/abs/1212.0830}{{\tt arXiv:1212.0830}}].

\bibitem{Fahy:2014jxa}
B.~Fahy, J.~Bulava, B.~{H\"orz}, K.~J. Juge, C.~Morningstar, {\em et.~al.},
  \href{http://xxx.lanl.gov/abs/1410.8843}{{\tt arXiv:1410.8843}}.

\bibitem{Metivet:2014bga}
{\bf BMW Collaboration}, T.~Metivet,
  \href{http://xxx.lanl.gov/abs/1410.8447}{{\tt arXiv:1410.8447}}.

\bibitem{Fu:2012tj}
Z.~Fu and K.~Fu, {\em Phys.Rev.} {\bf D86} (2012) 094507,
  [\href{http://xxx.lanl.gov/abs/1209.0350}{{\tt arXiv:1209.0350}}].

\bibitem{Prelovsek:2013ela}
S.~Prelovsek, L.~Leskovec, C.~Lang, and D.~Mohler, {\em Phys.Rev.} {\bf D88}
  (2013), no.~5 054508, [\href{http://xxx.lanl.gov/abs/1307.0736}{{\tt
  arXiv:1307.0736}}].

\bibitem{Dudek:2014qha}
{\bf Hadron Spectrum}, J.~J. Dudek, R.~G. Edwards, C.~E. Thomas, and D.~J.
  Wilson, {\em Phys.Rev.Lett.} {\bf 113} (2014), no.~18 182001,
  [\href{http://xxx.lanl.gov/abs/1406.4158}{{\tt arXiv:1406.4158}}].

\bibitem{Ozaki:2012ce}
S.~Ozaki and S.~Sasaki, {\em Phys.Rev.} {\bf D87} (2013) 014506,
  [\href{http://xxx.lanl.gov/abs/1211.5512}{{\tt arXiv:1211.5512}}].

\bibitem{Verduci:2014csa}
V.~Verduci and C.~B. Lang, \href{http://xxx.lanl.gov/abs/1412.0701}{{\tt
  arXiv:1412.0701}}.

\bibitem{Giedt:2014ysa}
J.~Giedt and D.~Howarth, \href{http://xxx.lanl.gov/abs/1405.4524}{{\tt
  arXiv:1405.4524}}.

\bibitem{Wakayama:2014gpa}
M.~Wakayama, T.~Kunihiro, S.~Muroya, A.~Nakamura, C.~Nonaka, {\em et.~al.},
  \href{http://xxx.lanl.gov/abs/1412.3909}{{\tt arXiv:1412.3909}}.

\bibitem{Abdel-Rehim:2014zwa}
A.~Abdel-Rehim, C.~Alexandrou, J.~Berlin, M.~D. Brida, M.~Gravina, {\em
  et.~al.}, \href{http://xxx.lanl.gov/abs/1410.8757}{{\tt arXiv:1410.8757}}.

\bibitem{Ishii:2006ec}
N.~Ishii, S.~Aoki, and T.~Hatsuda, {\em Phys. Rev. Lett.} {\bf 99} (2007)
  022001, [\href{http://xxx.lanl.gov/abs/nucl-th/0611096}{{\tt
  nucl-th/0611096}}].

\bibitem{Aoki:2009ji}
S.~Aoki, T.~Hatsuda, and N.~Ishii, {\em Prog.Theor.Phys.} {\bf 123} (2010)
  89--128, [\href{http://xxx.lanl.gov/abs/0909.5585}{{\tt arXiv:0909.5585}}].

\bibitem{Lin:2001ek}
C.~D. Lin, G.~Martinelli, C.~T. Sachrajda, and M.~Testa, {\em Nucl.Phys.} {\bf
  B619} (2001) 467--498, [\href{http://xxx.lanl.gov/abs/hep-lat/0104006}{{\tt
  hep-lat/0104006}}].

\bibitem{Balog:1999ww}
J.~Balog, M.~Niedermaier, F.~Niedermayer, A.~Patrascioiu, E.~Seiler, {\em
  et.~al.}, {\em Phys.Rev.} {\bf D60} (1999) 094508,
  [\href{http://xxx.lanl.gov/abs/hep-lat/9903036}{{\tt hep-lat/9903036}}].

\bibitem{Aoki:2005uf}
{\bf CP-PACS Collaboration}, S.~Aoki {\em et.~al.}, {\em Phys. Rev.} {\bf D71}
  (2005) 094504, [\href{http://xxx.lanl.gov/abs/hep-lat/0503025}{{\tt
  hep-lat/0503025}}].

\bibitem{HALQCD:2012aa}
{\bf HALQCD Collaboration}, N.~Ishii {\em et.~al.}, {\em Phys.Lett.} {\bf B712}
  (2012) 437--441, [\href{http://xxx.lanl.gov/abs/1203.3642}{{\tt
  arXiv:1203.3642}}].

\bibitem{Ishii:2013ira}
{\bf HALQCD Collaboration}, N.~Ishii, {\em PoS} {\bf CD12} (2013) 025.

\bibitem{Yamada:2014jra}
{\bf HALQCD Collaboration}, M.~Yamada, {\em PoS} {\bf LATTICE2013} (2014) 232.

\bibitem{Doi:2011gq}
{\bf HALQCD Collaboration}, T.~Doi {\em et.~al.}, {\em Prog.Theor.Phys.} {\bf
  127} (2012) 723--738, [\href{http://xxx.lanl.gov/abs/1106.2276}{{\tt
  arXiv:1106.2276}}].

\bibitem{Kurth:2013tua}
T.~Kurth, N.~Ishii, T.~Doi, S.~Aoki, and T.~Hatsuda, {\em JHEP} {\bf 1312}
  (2013) 015, [\href{http://xxx.lanl.gov/abs/1305.4462}{{\tt
  arXiv:1305.4462}}].

\bibitem{Inoue:2010es}
{\bf HALQCD Collaboration}, T.~Inoue {\em et.~al.}, {\em Phys.Rev.Lett.} {\bf
  106} (2011) 162002, [\href{http://xxx.lanl.gov/abs/1012.5928}{{\tt
  arXiv:1012.5928}}].

\bibitem{Inoue:2011ai}
{\bf HALQCD Collaboration}, T.~Inoue {\em et.~al.}, {\em Nucl. Phys.} {\bf
  A881} (2012) 28--43, [\href{http://xxx.lanl.gov/abs/1112.5926}{{\tt
  arXiv:1112.5926}}].

\bibitem{Green:2014dea}
J.~Green, A.~Francis, P.~Junnarkar, C.~Miao, T.~Rae, {\em et.~al.},
  \href{http://xxx.lanl.gov/abs/1411.1643}{{\tt arXiv:1411.1643}}.

\bibitem{Francis:2015bnl}
A.~Francis, J.~Green, P.~Junnarkar, C.~Miao, T.~Rae, and H.~Wittig.
\newblock
  https://indico.bnl.gov/getFile.py/access?contribId=28\&sessionId=9\&resId=0\&materialId=slides\&confId=934.

\bibitem{Fukugita:1994ve}
M.~Fukugita, Y.~Kuramashi, M.~Okawa, H.~Mino, and A.~Ukawa, {\em Phys. Rev.}
  {\bf D52} (1995) 3003--3023,
  [\href{http://xxx.lanl.gov/abs/hep-lat/9501024}{{\tt hep-lat/9501024}}].

\bibitem{Beane:2006mx}
S.~R. Beane, P.~F. Bedaque, K.~Orginos, and M.~J. Savage, {\em Phys. Rev.
  Lett.} {\bf 97} (2006) 012001,
  [\href{http://xxx.lanl.gov/abs/hep-lat/0602010}{{\tt hep-lat/0602010}}].

\bibitem{Beane:2011iw}
{\bf NPLQCD Collaboration}, S.~Beane {\em et.~al.}, {\em Phys. Rev.} {\bf D85}
  (2012) 054511, [\href{http://xxx.lanl.gov/abs/1109.2889}{{\tt
  arXiv:1109.2889}}].

\bibitem{Takahashi:2009ef}
T.~T. Takahashi and Y.~Kanada-En'yo, {\em Phys.Rev.} {\bf D82} (2010) 094506,
  [\href{http://xxx.lanl.gov/abs/0912.0691}{{\tt arXiv:0912.0691}}].

\bibitem{Murano:2011nz}
K.~Murano, N.~Ishii, S.~Aoki, and T.~Hatsuda, {\em Prog.Theor.Phys.} {\bf 125}
  (2011) 1225--1240, [\href{http://xxx.lanl.gov/abs/1103.0619}{{\tt
  arXiv:1103.0619}}].

\bibitem{Doi:2013haa}
{\bf HALQCD Collaboration}, T.~Doi, {\em PoS} {\bf LATTICE2013} (2014) 226,
  [\href{http://xxx.lanl.gov/abs/1311.2697}{{\tt arXiv:1311.2697}}].

\bibitem{Walker-Loud:2014iea}
A.~Walker-Loud, {\em PoS} {\bf LATTICE2013} (2014) 013,
  [\href{http://xxx.lanl.gov/abs/1401.8259}{{\tt arXiv:1401.8259}}].

\bibitem{Yamazaki:2009ua}
{\bf PACS-CS Collaboration}, T.~Yamazaki, Y.~Kuramashi, and A.~Ukawa, {\em
  Phys.Rev.} {\bf D81} (2010) 111504,
  [\href{http://xxx.lanl.gov/abs/0912.1383}{{\tt arXiv:0912.1383}}].

\bibitem{Doi:2012xd}
T.~Doi and M.~G. Endres, {\em Comput.Phys.Commun.} {\bf 184} (2013) 117,
  [\href{http://xxx.lanl.gov/abs/1205.0585}{{\tt arXiv:1205.0585}}].

\bibitem{Detmold:2012eu}
W.~Detmold and K.~Orginos, {\em Phys.Rev.} {\bf D87} (2013), no.~11 114512,
  [\href{http://xxx.lanl.gov/abs/1207.1452}{{\tt arXiv:1207.1452}}].

\bibitem{Gunther:2013xj}
J.~{G\"unther}, B.~C. Toth, and L.~Varnhorst, {\em Phys.Rev.} {\bf D87} (2013),
  no.~9 094513, [\href{http://xxx.lanl.gov/abs/1301.4895}{{\tt
  arXiv:1301.4895}}].

\bibitem{Vachaspati:2014bda}
P.~Vachaspati and W.~Detmold, \href{http://xxx.lanl.gov/abs/1411.3691}{{\tt
  arXiv:1411.3691}}.

\bibitem{Beane:2014ora}
{\bf NPLQCD Collaboration}, S.~Beane, E.~Chang, S.~Cohen, W.~Detmold, H.~Lin,
  {\em et.~al.}, {\em Phys.Rev.Lett.} {\bf 113} (2014), no.~25 252001,
  [\href{http://xxx.lanl.gov/abs/1409.3556}{{\tt arXiv:1409.3556}}].

\bibitem{Beane:2014sda}
{\bf NPLQCD Collaboration}, S.~Beane, E.~Chang, S.~Cohen, W.~Detmold, H.~W.
  Lin, {\em et.~al.}, \href{http://xxx.lanl.gov/abs/1410.7069}{{\tt
  arXiv:1410.7069}}.

\bibitem{Etminan:2014tya}
{\bf HALQCD Collaboration}, F.~Etminan {\em et.~al.}, {\em Nucl.Phys.} {\bf
  A928} (2014) 89--98, [\href{http://xxx.lanl.gov/abs/1403.7284}{{\tt
  arXiv:1403.7284}}].

\bibitem{Detmold:2014kba}
W.~Detmold, M.~McCullough, and A.~Pochinsky, {\em Phys.Rev.} {\bf D90} (2014),
  no.~11 114506, [\href{http://xxx.lanl.gov/abs/1406.4116}{{\tt
  arXiv:1406.4116}}].

\end{thebibliography}\endgroup

\end{document}